\documentclass[12pt]{article}
\usepackage{amsmath}
\usepackage{graphicx,psfrag,epsf}
\usepackage{enumerate}
\usepackage{natbib}
\usepackage{url} 
\usepackage{multirow}
\usepackage{booktabs}
\usepackage{multirow}
\usepackage{verbatim}
\usepackage{caption}
\usepackage{setspace}
\usepackage{authblk}
\newcommand{\var}{\mbox{Var}}
\newcommand{\bmu}{\mbox{\boldmath $\mu$}}
\newcommand{\bdelta}{\mbox{\boldmath $\delta$}}

\begin{document}

\def\spacingset#1{\renewcommand{\baselinestretch}%
{#1}\small\normalsize} \spacingset{1}


  \title{\bf Individualized Treatment Effects with Censored Data via Fully Nonparametric Bayesian Accelerated Failure Time Models}
  
\author[1]{Nicholas C. Henderson}
\author[2]{Thomas A. Louis}
\author[1,2]{Gary L. Rosner}
\author[1,2]{Ravi Varadhan}
\affil[1]{ {\small Sidney Kimmel Comprehensive Cancer Center, Johns Hopkins University } }
\affil[2]{ {\small Department of Biostatistics, Bloomberg School of Public Health, Johns Hopkins University} }

\date{\vspace{-6ex}}
  \maketitle

\bigskip
\begin{abstract}
Individuals often respond differently to identical treatments, and characterizing 
such variability in treatment response is an important aim in the practice of 
personalized medicine. In this article, we describe a non-parametric accelerated failure
time model that can be used to analyze heterogeneous treatment effects (HTE)
when patient outcomes are time-to-event. 
By utilizing Bayesian additive regression trees and a mean-constrained Dirichlet process 
mixture model, our approach offers a flexible model for the regression function
while placing few restrictions on the baseline hazard. 
Our non-parametric method leads to natural estimates of individual treatment effect and has the flexibility to address many major goals of HTE assessment. Moreover, our method requires little user input in terms of tuning parameter selection or subgroup specification. 
We illustrate the merits of our proposed approach with a detailed analysis of two large clinical trials for the prevention 
and treatment of congestive heart failure using an angiotensin-converting enzyme inhibitor.  
The analysis revealed considerable evidence for the presence of HTE in both trials
as demonstrated by substantial estimated variation in treatment effect
and by high proportions of patients exhibiting strong evidence of having
treatment effects which differ from the overall treatment effect.
\end{abstract}

\noindent%
\vfill

\newpage
\spacingset{1.4}

\vspace{-6pt}
\section{Introduction} \label{sec:intro}
\vspace{-3pt}

While the main focus of clinical trials is on evaluating the average effect of a particular treatment,
assessing heterogeneity in treatment effect (HTE) across key patient sub-populations remains an important
task in evaluating the results of clinical studies. Accurate evaluations of HTE that is attributable to variation in baseline patient characteristics offers many potential benefits in terms of informing patient decision-making  
and in appropriately targeting existing therapies.
HTE assessment can encompass a wide range of goals:  quantification of overall heterogeneity in treatment response,  identification of important patient characteristics related to HTE, estimation of proportion who benefits from the treatment, identification of patient sub-populations deriving most benefit from treatment, detection of cross-over (qualitative) interactions, identifying the patients who are harmed by treatment, estimation of individualized treatment effects, optimal treatment allocation for individuals, and predicting treatment effect for a future patient. 

Recently, there has been increasing methodology development in the arena of HTE assessment.  However, each developed method has been targeted to address one specific goal of HTE analysis. For example, \cite{xu:2015}, \cite{Foster:2011} and \cite{berger:2014} developed a method to identify patient subgroups whose response to treatment differs substantially from the average treatment effect. \cite{Weisberg:2015} and \cite{Lamont:2016} discuss estimation of individualized treatment effects. \cite{Zhao:2012} and \cite{zhao:2015} discuss construction of optimal individualized treatment rules
through minimization of a weighted classification error. \cite{shen:2016} focus on detection of biomarkers which are predictive of treatment effect heterogeneity. Thus, none of the existing methods is sufficiently flexible to address multiple goals of HTE analysis.  
  
Our aim in this paper is to construct a unified methodology for analyzing and exploring 
HTE with a particular focus on cases where the responses are time-to-event. 
The methodology is readily extended to continuous and binary response data. 
The motivation for investigating such a framework is the recognition that most, if not all,
of the above-stated goals of personalized medicine could be directly addressed if a sufficiently rich
approximation to the true data generating model for patient outcomes were available.
Bayesian nonparametric methods are well-suited to provide this more unified framework for HTE analysis
because they place few a priori restrictions on the form of the data-generating model and provide great adaptivity.
Bayesian nonparametrics allow construction of flexible models for patient outcomes coupled with probability modeling of all unknown quantities which generates a full posterior distribution over the desired response surface.
This allows researchers to directly address a wide range of inferential targets 
without the need to fit a series of separate models or to employ a series of different procedures. Our methodology has the flexibility to address all of the HTE goals previously highlighted. For example, the researchers could quantify overall HTE; identify most important patient characteristics pertaining to HTE; estimate the proportion benefiting from, or harmed by, the treatment; and predict treatment effect for a future patient.

Bayesian additive regression trees (BART) (\cite{chipman:2010}) provide a flexible means of modeling
patient outcomes without the need for making specific parametric assumptions,
specifying a functional form for a regression model, or for using pre-specified patient subgroups.
Because it relies on an ensemble of regression trees, BART has the capability to 
automatically detect non-linearities and covariate interactions.
As reported by \cite{Hill:2011} in the context of using BART for causal inference,
BART has the advantage of exhibiting strong predictive performance in a variety of settings while
requiring little user input in terms of selecting tuning parameters.
While tree-based methods have been employed in the context of personalized medicine and subgroup identification by a variety of 
investigators including, for example, \cite{Su:2009}, \cite{Loh:2015}, \cite{chen:2016}, and \cite{Foster:2011},
BART offers several advantages for the analysis of HTE. 
In contrast to many other tree-based procedures that use a more algorithmic approach,
BART is model-based and utilizes a full likelihood function and corresponding prior over the
tree-related parameters. Because of this, BART automatically generates measures of posterior uncertainty;
on the other hand, reporting uncertainty intervals is often quite challenging for other frequentist tree-based procedures.
In addition, because inference with BART relies on posterior sampling, analysis of HTE on alternative 
treatment scales can be done directly by simply transforming the desired parameters in posterior sampling. 
Moreover, any quantity of interest for individualized decisions or HTE evaluation can be readily accommodated
by the Bayesian framework. In this paper, we aim to utilize and incorporate these 
advantages of BART into our approach for analyzing HTE with censored data.

Extensions of the original BART procedure to handle time-to-event outcomes have been proposed
and investigated by \cite{bonato:2011} and \cite{sparapani:2016}. In \cite{bonato:2011},
the authors introduce several sum-of-trees models and examine their use in
utilizing gene expression measurements for survival prediction. 
Among the survival models proposed by \cite{bonato:2011} is an accelerated failure time (AFT)
model with a sum-of-trees regression function and a normally distributed residual term.
\cite{sparapani:2016} introduce a non-parametric approach that employs BART to directly model
the individual-specific probabilities of an event occurring at the observed event and censoring times.
In contrast to this approach, we propose a non-parametric version of the AFT model which combines
a sum-of-trees model with a Dirichlet process mixture model for the residual distribution.
Such an approach has the advantage of providing great flexibility
while generating interpretable measures of covariate-specific treatment effects thus
facilitating the analysis of HTE.

Accelerated failure time (AFT) models (\cite{louis:1981}, \cite{Robins:1992}, or \cite{Wei:1992})
represent an alternative 
to Cox-proportional hazards models in the analysis of time-to-event data.
AFT models have a number of features which make them appealing in the context of personalized medicine and investigating the 
comparative effectiveness of different treatments.
Because they involve a regression with log-failure times as the response variable,
AFT models provide a direct interpretation of the relationship between patient covariates
and failure times. Moreover, treatment effects may be defined directly in terms
of the underlying failure times. 
Bayesian semi-parametric approaches to the accelerated failure time model have been investigated 
by a number of authors including \cite{christensen:1988},
\cite{kuo:1997}, \cite{hanson:2002}, and \cite{hanson:2006}.
\cite{kuo:1997} assume a parametric model for the regression function and 
suggest either modeling the distribution of the residual term or of the exponential of the residual
term via a Dirichlet process mixture model, while \cite{hanson:2006} proposed modeling the residual
distribution with a Dirichlet process mixture of Gamma densities.
Our approach for modeling the residual distribution resembles that of \cite{kuo:1997}.
Similar to these approaches, we model the residual distribution as a location-mixture of Gaussian densities, and
by utilizing constrained Dirichlet processes, we constrain the mean of the residual distribution to be zero, thereby
clarifying the interpretation of the regression function.

This paper is organized as follows. In Section \ref{s:model_description}, we describe
the general structure of our nonparametric, tree-based accelerated failure time model,
discuss its use in estimating individualized treatment effects, and describe
our approach for posterior computation. 
Section \ref{s:hte_inference} examines
several key inferential targets in the analysis of heterogeneous treatment effects 
and how the nonparametric AFT model may be utilized to estimate these targets. 
In Section \ref{s:simulations}, we detail the results of several simulation studies that
evaluate our procedure in terms of individualized treatment effect estimation, coverage,
and optimal treatment assignment.
In Section \ref{s:hte_example}, we examine a clinical trial involving the use of an ACE inhibitor,
and we demonstrate the use of our nonparametric AFT method to investigate heterogeneity of treatment effect in this study.  
We conclude in Section \ref{s:discussion} with a few final remarks.
The methods described in this paper are implemented in the R package \textbf{AFTrees}, which is
available for download at \url{http://www.hteguru.com/software}.

\vspace{-6pt}
\section{The Model} \label{s:model_description}
\vspace{-3pt}
\subsection{Notation and Non-parametric AFT model} \label{ss:notation}
\vspace{-3pt}
We assume that study participants have been randomized to one of two treatments which
we denote by either $A = 0$ or $A = 1$. We let $\mathbf{x}$ denote a $p \times 1$ vector 
of baseline covariates and let $T$ denote the failure time. 
Given a censoring time $C$, we observe $Y = \min\{T, C\}$ and a failure
indicator $\delta = \mathbf{1}\{ T \leq C\}$.
The data consist of $n$ independent measurements $\{ (Y_{i}, \delta_{i}, A_{i}, \mathbf{x}_{i}); i=1,\ldots,n \}$.
Although we assume randomized treatment assignment here, our approach may certainly
be applied in observational settings. In such settings, however, 
one should ensure that appropriate unconfoundedness assumptions (e.g., \cite{Hill:2011})
are reasonable, so that the individualized treatment effects defined in
(\ref{eq:ITE}) correspond to an expected difference in potential outcomes
under the two treatments.

The conventional accelerated failure time (AFT) model assumes that log-failure times
are linearly related to patient covariates.
We consider here a non-parametric analogue of the AFT model in which the failure time $T$ is 
related to the covariates and treatment assignment through
\begin{equation}
\log T  = m(A, \mathbf{x} )  +  W, 
\label{eq:aft_regression}
\end{equation}
and where the distribution of the residual term $W$ is assumed to satisfy $E(W) = 0$.
With the mean-zero constraint on the residual distribution, 
the regression function $m(A, \mathbf{x} )$ has a direct interpretation as the expected log-failure time given
treatment assignment and baseline covariates.

The AFT model (\ref{eq:aft_regression}) leads to a natural, directly interpretable definition
of the individualized treatment effect (ITE), namely, the difference in expected log-failure
in treatment $A = 1$ versus $A = 0$. Specifically, we define the ITE $\theta( \mathbf{x} )$
for a patient with covariate vector $\mathbf{x}$ as
\begin{eqnarray}
\theta( \mathbf{x}) &=& E\{ \log(T)| A=1, \mathbf{x}, m \} - E\{ \log(T)| A=0, \mathbf{x}, m \} \nonumber \\
&=& m(1, \mathbf{x}) - m(0, \mathbf{x}).
\label{eq:ITE}
\end{eqnarray}
 
The distribution of $T$ in the accelerated failure time model (\ref{eq:aft_regression})
is characterized by both the regression function $m$ and
the distribution $F_{W}$ of the residual term. In the following, we outline
a model for the regression function that utilizes additive regression trees,
and we describe a flexible nonparametric mixture model for the residual distribution $F_{W}$. 
\vspace{-6pt}
\subsection{Overview of BART} \label{ss:bart_overview}
\vspace{-3pt}
Bayesian Additive Regression Trees (BART) is an ensemble method in which the regression function
is represented as the sum of individual regression trees.
The BART model for the regression function relies on a collection of $J$ binary trees 
$\{ \mathcal{T}_{1}, \ldots, \mathcal{T}_{J} \}$ and an associated 
set of terminal node values $B_{j} = \{ \mu_{j,1}, \ldots, \mu_{j,n_{j}} \}$ for each binary tree $\mathcal{T}_{j}$. 
Each tree $\mathcal{T}_{j}$ consists of a sequence of decision rules through which any covariate vector
can be assigned to one terminal node of $\mathcal{T}_{j}$ by following the decision rules prescribed at
each of the interior nodes. In other words, each binary tree generates a partition of the predictor space 
in which each element $\mathbf{u} = ( A,\mathbf{x} )$ of the predictor space belongs to exactly one of the $n_{j}$ terminal 
nodes of $\mathcal{T}_{j}$. 
The decision rules at the interior nodes of $\mathcal{T}_{j}$ are of the
form $\{u_{k} \leq c\}$ vs. $\{ u_{k} > c \}$, where $u_{k}$ denotes the $k^{th}$ element of $\mathbf{u}$. 
A covariate $\mathbf{u}$ that corresponds to the $l^{th}$ terminal node of $\mathcal{T}_{j}$ is assigned the value $\mu_{j,l}$
and $g(A, \mathbf{x}; \mathcal{T}_{j}, B_{j})$ is used to denote the function that returns $\mu_{j,l} \in B_{j}$
whenever $(A, \mathbf{x})$ is assigned to the $l^{th}$ terminal node of $\mathcal{T}_{j}$.

The regression function $m$ is represented in BART as a sum of the individual tree contributions
\begin{equation}
m(A, \mathbf{x} ) = \sum_{j=1}^{J} g(A, \mathbf{x}; \mathcal{T}_{j}, B_{j}).
\label{eq:sum_of_trees}
\end{equation} 
Trees $\mathcal{T}_{j}$ and node values $B_{j}$ can be thought of as model parameters with priors
on these parameters inducing a prior on the regression function $m$ via (\ref{eq:sum_of_trees}).
To complete the description of the prior on $(\mathcal{T}_{1},B_{1}), \ldots, 
(\mathcal{T}_{J}, B_{J})$, one needs to specify the following: 
(i) the distribution on the choice of splitting variable at each internal node;
(ii) the distribution of the splitting value $c$ used at each internal node;
(iii) the probability that a node at a given node-depth $d$ splits,
which is assumed to be equal to $\alpha(1 + d)^{-\beta}$; 
and (iv) the distribution of the terminal node values $\mu_{j,l}$ which 
is assumed to be $\mu_{j,l} \sim \textrm{Normal}\{ 0, (4k^{2}J)^{-1} \}$.
In order to ensure that the prior variance for $\mu_{j,l}$ induces a prior on the regression function
that assigns high probability to the observed range of the data,  
\cite{chipman:2010} center and scale the response so that
the minimum and maximum values of the transformed response are $-0.5$ and $0.5$ respectively. 
The distributions used for (i) and (ii) are discussed in \cite{chipman:1998} and \cite{chipman:2010}. 

To denote the distribution on the regression function $m$
induced by the prior distribution on $\mathcal{T}_{j}, B_{j}$
with parameter values $(\alpha, \beta, k)$ and $J$ total trees,
we use the notation
\begin{equation}
m \sim \textrm{BART}(\alpha, \beta, k, J).
\nonumber
\end{equation} 
\vspace{-6pt}
\subsection{Centered DP mixture prior} \label{ss:cdp}
\vspace{-3pt}
We model the density $f_{W}$ of $W$ as a location-mixture of Gaussian densities
with common scale parameter $\sigma$. Letting $G$ denote the distribution of the locations,
we assume the density of $W$ (conditional on $G$ and $\sigma)$ can be expressed as
\begin{equation}
f_{W}(w| G, \sigma) = \frac{1}{\sigma} \int \phi\Big( \frac{w - \tau}{\sigma} \Big) dG(\tau),
\label{eq:w_mixturemodel}
\end{equation}
where $\phi(\cdot)$ is the standard normal density function. 
The Dirichlet process (DP) is a widely used choice for a nonparametric prior on an unknown probability distribution,
and the resulting DP mixture model for the distribution of $W$ 
provides a flexible prior for the residual density. 
Indeed, a DP mixture model similar to (\ref{eq:w_mixturemodel}) was used by \cite{kuo:1997}
as a prior for a smooth residual distribution in a semi-parametric 
accelerated failure time model.

Because of the zero-mean constraint on the residual distribution, 
the Dirichlet process is not an appropriate choice for a prior on $G$.  
A direct approach proposed by \cite{Dunson:2010} addresses the problem of placing mean and variance 
constraints on an unknown probability measure by utilizing a parameter-expanded
version of the Dirichlet process which the authors refer to as the centered Dirichlet process (CDP).
As formulated by \cite{Dunson:2010}, the CDP with mass parameter $M$ and base measure $G_{0}$ 
has the following stick-breaking representation
\begin{eqnarray}
G &=& \sum_{h=1}^{\infty} \pi_{h} \delta_{\tau_{h}} \nonumber \\ 
\pi_{h} &=& V_{h}\prod_{l < h} (1 - V_{l}), \quad h=1,\ldots,\infty  \nonumber \\
\tau_{h} &=& \tau_{h}^{*} - \mu_{G^{*}}, \quad h=1,\ldots,\infty \nonumber \\
V_{h} &\sim& \textrm{Beta}(1, M), \nonumber \\
\tau_{h}^{*} &\sim& G_{0}, \quad h=1,\ldots,\infty,   
\label{eq:cdp_stick}
\end{eqnarray}
where $\mu_{G^{*}} = \sum_{h=1}^{\infty} \pi_{h} \tau_{h}^{*}$
and where $\delta_{\tau}$ denotes a distribution that only consists of a point mass at $\tau$.
We denote that a random measure $G$ follows a centered Dirichlet process
with the notation $G \sim \textrm{CDP}(M, G_{0})$. From the above representation of the CDP, it is clear that the
mixture model (\ref{eq:w_mixturemodel}) for $W$ and the assumption that $G \sim \textrm{CDP}(M, G_{0})$ together imply the 
mean-zero constraint, since the expectation of $W$ can then be expressed as
\begin{equation}
E(W|G,\sigma) = \sum_{h=1}^{\infty} \tau_{h}\pi_{h}
= \sum_{h=1}^{\infty} \tau_{h}^{*}\pi_{h} - \mu_{G^{*}}\sum_{h=1}^{\infty} \pi_{h},  \nonumber
\end{equation}
which equals zero almost surely. 

For the scale parameter of $f_{W}$,
we assume that $\sigma^{2}$ follows an inverse chi-square distribution, $\sigma^{2} \sim \kappa\nu/\chi_{\nu}^{2}$,
with the default degrees of freedom $\nu$ set to $\nu = 3$. 
Instead of specifying a particular value for the mass parameter, we allow for learning about this parameter by 
assuming $M \sim \textrm{Gamma}(\psi_{1}, \psi_{2})$ where $\psi_{1}$ and $\psi_{2}$
refer to the shape and rate parameters of Gamma distribution respectively.

Our non-parametric model that combines the BART model for
the regression function and DP mixture model for the residual density can now 
be expressed hierarchically as
\begin{eqnarray}
\log T_{i} = m(A_{i}, \mathbf{x}_{i}) + W_{i}, \quad
W_{i}|\tau_{i},\sigma^{2} \sim N(\tau_{i}, \sigma^{2}), \textrm{  for } i=1, \ldots, n \nonumber \\
m \sim \textrm{BART}(\alpha, \beta, k, J), \quad \tau_{i}|G \sim G, \quad  G|M \sim \textrm{CDP}(M, G_{0}) \nonumber \\
\sigma^{2} \sim \kappa\nu/\chi_{\nu}^{2}, \quad
M \sim \textrm{Gamma}( \psi_{1}, \psi_{2}).
\label{eq:aft_hierarchy}
\end{eqnarray}
In our implementation, the base measure $G_{0}$ is assumed to be Gaussian with mean zero and variance $\sigma_{\tau}^{2}$.
Choosing $G_{0}$ to be conjugate to the Normal distribution simplifies posterior computation considerably, but
other choices of $G_{0}$ could be considered. For example, a t-distributed base measure could be implemented by introducing
an additional latent scale parameter.
\vspace{-6pt}
\subsection{Prior Specification}
\vspace{-3pt}
\textit{Prior for Trees and Terminal Node Parameters.} 
For the hyperparameters of the trees $\mathcal{T}_{1},\ldots,\mathcal{T}_{J}$, 
we defer to the defaults suggested in \cite{chipman:2010}; namely, $\alpha = 0.95$, $\beta = 2$, and $J = 200$.
These default settings seem to work quite well in practice, and in Section \ref{s:hte_example} we investigate the impact of varying
$J$ through cross-validation estimates of prediction performance. 

As discussed in Section \ref{ss:bart_overview}, the original description of BART in \cite{chipman:2010}
employs a transformation of the response variable and sets the hyperparameter $k$ to $k = 2$
so that the regression function is assigned
substantial prior probability to the observed range of the response.
Because our responses $Y_{i}$ are right-censored,
we propose an alternative approach to transforming the responses and to setting the prior variance of the
terminal node parameters. 
Our suggested approach is to first fit a parametric AFT model that only has an intercept in the model 
and that assumes log-normal residuals.
This produces estimates of the intercept $\hat{\mu}_{AFT}$ and the residual scale $\hat{\sigma}_{AFT}$
which allows us to define the transformed ``centered" responses as
$y_{i}^{tr} = y_{i}\exp\{ -\hat{\mu}_{AFT} \}$. 
Turning to the prior variance of the terminal node parameters $\mu_{j,l}$, 
we assign the terminal node values $\mu_{j,l}$ the prior distribution 
$\mu_{j,l} \sim \textrm{Normal}\{0, \zeta^{2}/(4Jk^{2}) \}$, where $\zeta = 4\hat{\sigma}_{AFT}$.
This prior on $\mu_{j,l}$ induces a $\textrm{Normal}\{ 0, 4\hat{\sigma}_{AFT}^{2}/k^{2} \}$ prior on the regression 
function $m(A, \mathbf{x})$ and hence assigns approximately $95\%$ prior probability to the 
interval $[-4k^{-1}\hat{\sigma}_{AFT}, 4k^{-1}\hat{\sigma}_{AFT}] $.
Thus, the default setting of $k=2$ assigns $95\%$ prior probability to the interval $[-2\hat{\sigma}_{AFT}, 2\hat{\sigma}_{AFT}]$.
Note that assigning most of the prior probability to the interval $[-2\hat{\sigma}_{AFT}, 2\hat{\sigma}_{AFT}]$ is sensible
because this corresponds to the regression function for the ``centered" responses $y_{i}^{tr}$
rather than the original responses.

As described in \cite{chipman:1998} and \cite{chipman:2010}, the prior on the splitting values $c$ used at
each internal node is uniform over the finite set of available splitting values for the chosen splitting variable.  
In implementations of BART, the number of possible available splitting values is typically truncated
so that it cannot exceed a pre-specified maximum value. The default setting used in 
the \verb"BayesTree" package (\cite{btree_package}) has a maximum of $100$ possible split points 
for each covariate, and the default is to assign a uniform prior over
potential split points that are equally spaced over the range of the covariate. An alternative option offered 
in \verb"BayesTree" is
to, for each covariate, assign a uniform prior over the observed quantiles of the covariate
rather than the uniform prior over the observed range of the covariate.
Our default choice is to use the uniform prior over covariate quantiles for the split point prior rather
than the uniform prior over equally spaced points.
With this quantile-based prior, we found, in many simulations, improved performance in terms of coverage.

\textit{Residual Distribution Prior.}
Under the assumed prior for the mass parameter, we have $E[M|\psi_{1},\psi_{2}] = \psi_{1}/\psi_{2}$ and
$\var(M|\psi_{1},\psi_{2}) = \psi_{1}/\psi_{2}^{2}$. We set $\psi_{1} = 2$ and $\psi_{2} = 0.1$
so that the resulting prior on $M$ is relatively diffuse with $E[M|\psi_{1},\psi_{2}] = 20$,
$\var[M|\psi_{1},\psi_{2}] = 200$, and a prior mode of $10$.

When setting the defaults for the remaining hyperparameters $\kappa$ and $\sigma_{\tau}^{2}$, we adopt a similar strategy
to that used by \cite{chipman:2010} for BART when calibrating the prior for the residual variance.
There, they rely on a preliminary, rough overestimate $\hat{\sigma}^{2}$ of the residual variance parameter $\sigma^{2}$ 
and define the prior for $\sigma^{2}$ in such a way that there is $1 - q$ prior
probability that $\sigma^{2}$ is greater than the rough estimate $\hat{\sigma}^{2}$.
Here, $q$ may be regarded as an additional hyperparameter with the value of $q$
determining how conservative the prior of $\sigma^{2}$ is relative to the initial estimate
of the residual variance.
\cite{chipman:2010} suggest using $q = 0.90$ as the default whenever $\nu$ is set to $\nu = 3$.

Similar to the approach described above, we begin with a rough over-estimate $\hat{\sigma}_{W}^{2}$ of the variance of $W$
to calibrate our choices of $\kappa$ and $\sigma_{\tau}^{2}$.   
A direct way of generating the estimate $\hat{\sigma}_{W}^{2}$ is to 
fit a parametric AFT model with log-normal residuals and use the resulting estimate of the residual variance,
but other estimates could potentially be used.  
To connect the estimate $\hat{\sigma}_{W}^{2}$ with the hyperparameters $\kappa$ 
and $\sigma_{\tau}^{2}$ described in (\ref{eq:aft_hierarchy}), it is helpful to first note
that the conditional variance of the residual term can be expressed as
\begin{equation}
\var(W|G,\sigma) = \sigma^{2} + \sigma_{\tau}^{2}
\sum_{h=1}^{\infty} \frac{\pi_{h}}{\sigma_{\tau}^{2}}(\tau_{h}^{*} - \mu_{G^{*}})^{2}.
\label{eq:var_dpmix}
\end{equation}
Our aim then is to select $\kappa$ and $\sigma_{\tau}^{2}$ so that the induced prior
on the variance of $W$ assigns approximately $1 - q$ probability to the event $\big\{ \var(W|G,\sigma) > \hat{\sigma}_{W}^{2} \big\}$,
where $\hat{\sigma}_{W}^{2}$ is treated here as a fixed quantity.
As an approximation to the distribution of (\ref{eq:var_dpmix}), we use the approximation that
$\sum_{h=1}^{\infty} \frac{\pi_{h}}{\sigma_{\tau}^{2}}(\tau_{h}^{*} - \mu_{G^{*}})^{2}$
has a $\textrm{Normal}\{ 1, 2/(M+1) \}$ distribution (see \cite{Yamato:1984} or 
Appendix A for further details about this approximation). Assuming further
that $\kappa = \sigma_{\tau}^{2}$, we have that the variance of $W$ is approximately distributed
as $\sigma_{\tau}^{2}[ \nu/\chi_{\nu}^{2} + N(1,\{2(M+1)\}^{-1}) ]$ where $M \sim \textrm{Gamma}(\psi_{1},\psi_{2})$, 
and with this approximation, we can directly find a value of $\sigma_{\tau}^{2} = \kappa$ such that
$P\big\{  \var(W|G,\sigma) \leq \hat{\sigma}_{W}^{2} \big\} = q$. In 
contrast to the $q=0.9$ setting suggested in \cite{chipman:2010},
we set the default to $q = 0.5$. 
\vspace{-6pt}
\subsection{Posterior Computation}
\vspace{-3pt}
The original Gibbs sampler proposed in \cite{chipman:2010}
works by sequentially updating each tree while holding all other $J-1$ trees fixed. 
As a result, each iteration of the Gibbs sampler consists of $2J + 1$ steps where
the first $2J$ steps involve updating either one of the trees $T_{j}$ or terminal node
parameters $M_{j}$ and the last step involves updating the residual variance parameter. 
The Metropolis-Hastings algorithm used to update the individual trees
is discussed in \cite{chipman:1998}.
Our strategy for posterior computation is a direct extension of the original Gibbs
sampler, viz., after updating trees and terminal node parameters, we update the
parameters related to the residual distribution. Censored values are handled 
through a data augmentation approach where unobserved survival times
are imputed in each Gibbs iteration.

To sample from the posterior of the CDP, we adopt the blocked Gibbs sampling 
approach described in \cite{Ishwaran:2001}.
In this approach, the mixing distribution $G$ is truncated so that it only has
a finite number of components $H$ which is done by assuming that, $V_{h} \sim \textrm{Beta}(1,M)$ for
$h=1,\ldots,H-1$ and $V_{H} = 1$. This modification of the stick-breaking weights 
ensures that $\sum_{h=1}^{H} \pi_{h} = 1$. 
One advantage of using the truncation approximation is that it
makes posterior inferences regarding $G$ straightforward.
Additionally, when truncating the stick-breaking distribution,
using the CDP prior as opposed to a DP prior does not present any additional challenges for posterior computation
because the unconstrained parameters $\tau_{h}^{*}, \mu_{G^{*}}$ in (\ref{eq:cdp_stick}) may be updated 
as described in \cite{Ishwaran:2001} 
with the parameters of interest $\tau_{h}$ then being updated through the simple transformation
$\tau_{h} = \tau_{h}^{*} - \mu_{G^{*}}$. The upper bound on the number of components $H$
should be chosen to be relatively large (as a default, we set $H=50$), 
and in the Gibbs sampler, the maximum index of the occupied clusters should be monitored.
If a maximum index equal to $H$ occurs frequently in posterior sampling, $H$ should be increased.

In the description of the Gibbs sampler, we use $z_{i}$ to denote a latent variable 
that represents a transformed, imputed survival time, and we let $y_{i}^{c}$
denote the ``complete-data" survival times for the transformed survival times. 
That is, $y_{i}^{c,tr} = y_{i}^{tr}$ if $\delta_{i} = 1$ and
$y_{i}^{c,tr} = z_{i}$ if $\delta_{i} = 0$. For posterior computation related to 
the Dirichlet process mixture, we let $S_{i}$ denote the cluster to which
the $i^{th}$ observation has been assigned.
An outline of the steps involved in one iteration of our Gibbs sampler is provided below. 
\begin{itemize}\itemsep2pt
\item[1.]
Update trees $\mathcal{T}_{1},\ldots,\mathcal{T}_{J}$ and node parameters $B_{1},\ldots,B_{J}$ using
the Bayesian backfitting approach of \cite{chipman:2010} with $\log y_{i}^{c,tr} - \tau_{S_{i}}$ as the responses. 
Using the updated $\mathcal{T}_{1},\ldots,\mathcal{T}_{J}$ and $B_{1},\ldots,B_{J}$,
update $m(A_{i}, \mathbf{x}_{i})$, $i = 1, \ldots, n$.
\item[2.]
Update cluster labels $S_{1},\ldots,S_{n}$ by sampling with probabilities
\begin{equation}
P( S_{i}=h ) \propto \pi_{h}\phi\Big( \frac{\log y_{i}^{c,tr} - m(A_{i}, \mathbf{x}_{i}) 
- \tau_{h}}{\sigma} \Big), \nonumber
\end{equation}
and tabulate cluster membership counts $n_{h} = \sum_{i} \mathbf{1}\{ S_{i} = h \}$.
\item[3.]
Sample stick-breaking weights $V_{h}$, $h=1,\ldots,H-1$ as
$V_{h} \sim \textrm{Beta}(\alpha_{h}, \beta_{h})$ where
$\alpha_{h} = 1 + n_{h}$ and $\beta_{h} = M + \sum_{k=h+1}^{H} n_{k}$. Set $V_{H} = 1$.
The updated mixture proportions are then determined by $\pi_{h} = V_{h}\prod_{k < h}(1 - V_{k})$,
for $h = 1, \ldots, H$.
\item[4.]
Sample unconstrained cluster locations $\tau_{h}^{*}$ 
\begin{equation}
\tau_{h}^{*} \sim \textrm{Normal}\Big(
\frac{\sigma_{\tau}^{2}}{n_{h}\sigma_{\tau}^{2} + \sigma^{2}}
\sum_{i=1}^{n} \{ \log y_{i}^{c,tr} - m(A_{i}, \mathbf{x}_{i}) \}\mathbf{1}\{S_{i}=h\} ,
\frac{\sigma_{\tau}^{2}\sigma^{2}}{n_{h}\sigma_{\tau}^{2} + \sigma^{2} } \Big), \nonumber
\end{equation}
and update constrained cluster locations $\tau_{h} = \tau_{h}^{*} - \mu_{G^{*}}$,
where $\mu_{G^{*}} = \sum_{h=1}^{H} \pi_{h}\tau_{h}^{*}$.
\item[5.]
Update mass parameter $M \sim \textrm{Gamma}\Big( \psi_{1} + H - 1, \psi_{2} - \sum_{h=1}^{H-1}\log(1 - V_{h}) \Big)$
and scale parameter
$\sigma^{2} \sim \textrm{Inverse-Gamma}\big( \frac{\nu + n}{2}, \frac{\hat{s}^{2} + \kappa\nu}{2} \big)$,
where  $\hat{s}^{2}$ is given by
\begin{equation}
\hat{s}^{2} = \sum_{h = 1}^{H} \sum_{i = 1}^{n} \{ \log(y_{i}^{c,tr}) - m(A_{i},\mathbf{x}_{i}) - \tau_{h}  \}^{2}\mathbf{1}\{ S_{i} = h \}.
\nonumber
\end{equation}
\item[6.]
For each $i \in \{k: \delta_{k} = 0\}$, update $z_{i}$ by sampling
\begin{equation}
\log z_{i} \sim \textrm{Truncated-Normal}( m(A_{i}, \mathbf{x}_{i} ) + \tau_{S_{i}}, \sigma^{2}; \log y_{i}^{tr} ), \nonumber
\nonumber
\end{equation}
and set $y_{i}^{c,tr} = z_{i}$. Here, $X \sim \textrm{Truncated-Normal}(\mu, \sigma^{2}; a)$ means that
$X$ is distributed as $Z|Z > a$ where $Z \sim \textrm{Normal}(\mu, \sigma^{2})$.
\end{itemize}
Because we use the transformed responses 
$\log(y_{i}^{tr}) = \log(y_{i}) - \hat{\mu}_{AFT}$
in posterior computation, we add $\hat{\mu}_{AFT}$ to the posterior draws of $m(A,\mathbf{x})$ in the final output.

\vspace{-6pt}
\section{Posterior Inferences for the Analysis of Heterogeneous Treatment Effects} \label{s:hte_inference}
\vspace{-3pt}
The nonparametric AFT model (\ref{eq:aft_hierarchy}) generates a full posterior over the entire
regression function $m(A, \mathbf{x} )$ and the residual distribution. As such, 
this model has the flexibility to address a variety of questions related to heterogeneity of treatment effect. 
In particular, overall variation in response to treatment, proportion of patients likely to benefit from treatment
along with individual-specific treatment effects and survival curves 
may all be defined in terms of parameters from the nonparametric AFT model. 
\vspace{-6pt}
\subsection{Individualized Treatment Effects}
\vspace{-3pt}
As discussed in Section \ref{ss:notation}, a natural definition of the individual treatment effects in the context of an AFT model
is the difference in expected log-survival $\theta(\mathbf{x}) = m(1, \mathbf{x}) - m(0, \mathbf{x})$. 
Draws from the posterior distribution of $( m(A_{1},\mathbf{x}_{1}), \ldots, m(A_{n}, \mathbf{x}_{n}) )$ 
allow one to compute fully nonparametric estimates $\hat{\theta}( \mathbf{x}_{i} )$ of the treatment effects along with
corresponding $95 \%$ credible intervals.  
As is natural with an AFT model, the treatment difference $\theta( \mathbf{x} )$ in (\ref{eq:ITE}) 
is examined on the scale of log-survival time,
but other, more interpretable scales on which to report treatment effects could 
be easily computed. For example, ratios in expected survival times $\xi(\mathbf{x})$
defined by
\begin{equation}
\xi( \mathbf{x} ) = E\big\{ T| A = 1, \mathbf{x}, m \big\}/E\big\{ T | A = 0, \mathbf{x}, m \big\}
= \exp\{ \theta(\mathbf{x} ) \}  \nonumber
\end{equation}
could be estimated via posterior output.
Likewise, one could estimate differences in expected failure time by using both posterior draws of $\theta(\mathbf{x})$
and of the residual distribution.
Posterior information regarding treatment effects may be used to stratify patients 
into different groups based on anticipated treatment benefit. Stratification could be 
done using the posterior mean, the posterior probability of treatment benefit, or
some other relevant measure.  
\vspace{-6pt}
\subsection{Assessing Evidence Heterogeneity of Treatment Effect}
\vspace{-3pt}
As a way of detecting the presence of HTE, we examine the posterior probabilities of differential treatment effect
\begin{equation}
D_{i} = P\big\{ \theta(\mathbf{x}_{i}) \geq \bar{\theta}  | \mathbf{y}, \bdelta \big \}, 
\label{eq:pp_dte}
\end{equation}
along with closely related quantity
\begin{equation}
D_{i}^{*} = \max\{ 1 - 2D_{i}, 2D_{i} - 1\}, 
\label{eq:pp_dte_star}
\end{equation}
where, in (\ref{eq:pp_dte}), $\bar{\theta} = n^{-1}\sum_{i=1}^{n} \theta(\mathbf{x}_{i})$
is the conditional average treatment effect. Note that $\bar{\theta}$ is a model parameter that
represents the average value of the individual $\theta(\mathbf{x}_{i})$ and does not represent a posterior mean.
The posterior probability
$D_{i}$ is a measure of the evidence that the ITE $\theta(\mathbf{x}_{i})$ is greater or equal
to $\bar{\theta}$, 
and thus we should expect both high and low values of $D_{i}$ in settings where substantial HTE is present.
Note that $D_{i}^{*}$ approaches $1$ as the value of $D_{i}$ approaches either $0$ or $1$, and
$D_{i}^{*} = 0$ whenever $D_{i} = 1/2$.
For a given individual $i$, we consider there to be strong evidence of a differential treatment
effect if $D_{i}^{*} > 0.95$ (equivalently, if $D_{i} \leq 0.025$ or $D_{i} \geq 0.975$), and 
we define an individual as having mild evidence of a differential treatment effect 
provided that $D_{i}^{*} > 0.8$ (equivalently, if $D_{i} < 0.1$ or $D_{i} > 0.9$).
For cases with no HTE present,
the proportion of patients exhibiting strong evidence of differential treatment effect
should, ideally, be zero or quite close to zero. 
For this reason, the proportion of patients with $D_{i}^{*} > 0.95$
can potentially be a useful summary measure for detecting the presence of HTE.
In this paper, we do not explore explicit choices of a threshold for this proportion,
but we examine, through a simulation study in Section \ref{ss:null_sims},
the value of this proportion for scenarios for which no HTE is present
and later compare these simulation results with the observed proportion
of patients in the SOLVD trials who exhibit strong evidence of differential treatment effect.
It is worth mentioning that the quantity $D_{i}^{*}$ represents evidence that the treatment effect for patient $i$
differs from the overall treatment effect, and by itself, is not a robust indicator of HTE across patients in the trial.
Rather, the \textit{proportion} of patients with high values of $D_{i}^{*}$ 
is what we use as a means of assessing evidence for HTE.

It is worth noting that the presence or absence of HTE depends
on the treatment effect scale, and $D_{i}$ is designed for cases,
such as the AFT model, where HTE is difference in expected log-failure time. 
For example, it is possible to have heterogeneity on the log-hazard ratio
scale while having zero differences in the ITEs $\theta(\mathbf{x}_{i})$ across patients.
\vspace{-12pt}
\subsection{Characterizing Heterogeneity of Treatment Effect}
\vspace{-6pt}
Variability in treatment effect across patients in the study is a prime target of interest when evaluating
the extent of heterogeneous treatment effects from the results of a clinical trial.
Assessments of HTE can be used to evaluate consistency 
of response to treatment across patient sub-populations or to assess whether or not 
there are patient subgroups that appear to respond especially strongly to treatment.
In more conventional subgroup analyses (e.g., \cite{jones:2011}), heterogeneity in treatment effect is frequently reported in terms
of the posterior variation in treatment effect across patient subgroups. While the variance of treatment effect is a useful measure,
especially in the context of subgroup analysis,
we can provide a more detailed view of HTE by examining the full distribution 
of the individualized treatment effects defined by (\ref{eq:ITE}) where the distribution
may be captured by
the latent empirical distribution function $H_{n}(t) = \frac{1}{n} \sum_{i=1}^{n} \mathbf{1}\{ \theta( \mathbf{x}_{i} ) \leq t \}$.
Such an approach to examining the ``distribution" of a large collection of parameters has been explored in 
\cite{shen:1998} and \cite{louis:1999}. 
The distribution function $H_{n}(t)$ may be regarded as a model parameter 
that can directly estimated by
\begin{equation}
\hat{H}_{n}(t) = \frac{1}{n} \sum_{i=1}^{n} P\{ \theta( \mathbf{x}_{i} ) \leq t| \mathbf{y}, \bdelta \},
\label{eq:histogram_estimate}
\end{equation}
and credible bands for $H_{n}(t)$ may be obtained from posterior samples.  
For improved visualization of the spread of treatment effects, it is 
often better to display a density function $\hat{h}_{n}(t)$ associated with (\ref{eq:histogram_estimate})
which could be obtained through direct differentiation of (\ref{eq:histogram_estimate}). Alternatively, a 
smooth estimate can be found by computing the posterior mean of a kernel function $K_{\lambda}$ with
bandwidth $\lambda$ 
\begin{equation}
\hat{h}_{n}(t) = \frac{1}{n} \sum_{i=1}^{n} E\Big\{ K_{\lambda}\big( t - \theta(\mathbf{x}_{i})  \big) \Big| 
\mathbf{y}, \bdelta \Big\}.
\label{eq:smooth_density}
\end{equation}
The posterior of $H_{n}(t)$ provides a direct assessment of
the variation in the underlying treatment effects, and as such,
serves as a useful overall evaluation of HTE. 
\vspace{-6pt}
\subsection{Proportion Who Benefits}
\vspace{-3pt}
Another quantity of interest related to HTE is
the proportion of patients who benefit from treatment. 
Such a measure has a direct interpretation and is also a useful quantity for assessing
the presence of cross-over or qualitative interactions, namely, cases where the effect of treatment
has the opposite sign as the overall average treatment effect. 
That is, for situations where an overall treatment benefit has been determined, 
a low estimated proportion of patients benefiting may be an
indication of the existence of cross-over interactions.

Using the treatment differences $\theta(\mathbf{x})$, the proportion who benefit may be defined as
\begin{equation}
Q = \frac{1}{n}\sum_{i=1}^{n} \mathbf{1}\{ \theta( \mathbf{x}_{i} ) > 0\}.
\label{eq:prop_benefit}
\end{equation}
Alternatively, one could define the proportion benefiting relative to a clinically relevant threshold $\varepsilon > 0$,
i.e., $Q_{\epsilon} = n^{-1}\sum_{i=1}^{n} \mathbf{1}\{ \theta(\mathbf{x}_{i}) > \varepsilon \}$. 
The posterior mean of $Q$ is an average over patients 
of the posterior probabilities of treatment benefit $\hat{p}_{i} = P\{ \theta( \mathbf{x}_{i}) > 0|\mathbf{y}, \bdelta\}$. 
Posterior probabilities of treatment benefit can be used for 
treatment assignment ($\hat{p}_{i} > 1/2$ vs. $\hat{p}_{i} \leq 1/2$), 
or as an additional summary measure of HTE where one, for example, could tabulate 
the proportion very likely to benefit from treatment $\hat{p}_{i} > 0.99$ or
the proportion likely to benefit from treatment $\hat{p}_{i} > 0.90$.
\vspace{-6pt}
\subsection{Treatment Allocation}
\vspace{-3pt}
Individualized treatment recommendations may be directly obtained by combining a fit of the
non-parametric AFT model with a procedure minimizing the posterior risk associated
with a chosen loss function. For instance, when trying to minimize the proportion
of treatment misclassifications, 
one would assign treatment based on whether or not the posterior probability of the event 
$\{\theta(\mathbf{x}) > 0 \}$ was greater than $0.5$. 
Alternatively, one could optimize a weighted mis-classification loss where mis-classifications
are weighted by the corresponding magnitude $|\theta(\mathbf{x})|$ of the treatment effect,
in which case the optimal treatment decision would depend on the posterior mean
of $\mathbf{1}\{ \theta(\mathbf{x}) > 0 \} \times | \theta(\mathbf{x})|$. 
Though we do not explore the issue in this paper, such approaches to inividualized
treatment allocation could potentially be used, for example, in the development of
adaptive randomization strategies for clinical trials.
\vspace{-6pt}
\subsection{Individual-level Survival Functions}
\vspace{-3pt}
In terms of the quantities of the non-parametric AFT model (\ref{eq:aft_hierarchy}), individual-specific survival curves are defined by
\begin{equation}
P\{ T > t| A, \mathbf{x}, m, G, \sigma  \}
= 1 - \int \Phi \Big(  \frac{\log t - m(A, \mathbf{x}) - \tau}{\sigma}    \Big) dG(\tau). \nonumber 
\end{equation}
Using the truncated distribution $G_{H}$ as an approximation in posterior computation,
the survival curves are given by
\begin{equation}
P\{ T > t| A, \mathbf{x}, m, G_{H}, \sigma  \}
= 1 - \sum_{h=1}^{H} \Phi \Big(  \frac{\log t - m(A, \mathbf{x}) - \tau_{h}}{\sigma}    \Big) \pi_{h}, 
\label{eq:survival_formula}
\end{equation}
which may be directly estimated using posterior draws of the regression function and $\tau_{h}, \pi_{h}$.
\vspace{-6pt}
\subsection{Partial Dependence and Variable Importance}
\vspace{-3pt}
Partial dependence plots are a useful tool for visually assessing the dependence of 
an estimated function on a particular covariate or set of covariates. 
As described in \cite{friedman:2001}, such plots demonstrate the way an estimated function changes as a particular covariate
varies while averaging over the remaining covariates.

For the purposes of examining the impact of a covariate on the treatment effects, we define 
the partial dependence function for the $l^{th}$ covariate as
\begin{equation}
\rho_{l}( z ) = \frac{1}{n}\sum_{i=1}^{n} \theta( z, \mathbf{x}_{i,-l} ),
\nonumber
\end{equation}
where $(z,\mathbf{x}_{i,-l})$ denotes a vector where the $l^{th}$ component of $\mathbf{x}_{i}$ has been removed
and replaced with the value $z$. Estimated partial dependence functions $\hat{\rho}_{l}(z)$ with
associated credible bands may be obtained directly from MCMC output.

\vspace{-6pt}
\section{Simulations}  \label{s:simulations}
\vspace{-3pt}
To evaluate the performance of the non-parametric, tree-based AFT method, we performed 
three simulation studies. For performance related to quantifying HTE,
we recorded the following measures: root mean-squared error of the estimated individualized
treatment effects, the proportion of patients allocated to the wrong treatment,
and the average coverage of the confidence/credible intervals.
For each simulation, the root mean-squared error (RMSE) is measured as
$\sqrt{ n^{-1}\sum_{i=1}^{n} \{ \hat{\theta}(\mathbf{x}_{i}) - \theta(\mathbf{x}_{i}) \}^{2}} $,
where $\theta(\mathbf{x}_{i})$ is as defined in Section \ref{ss:notation}, and 
$\hat{\theta}(\mathbf{x}_{i})$ is an estimate of the ITE.
For treatment classification indicators $R_{1},\ldots,R_{n}$ (i.e., $R_{i} = 1$ if patient $i$ is
classified as benefiting from treatment $A=1$ rather than $A=0)$,
the proportion mis-classified (MCprop) is calculated within each simulation replication as
\begin{equation}
\textrm{MCprop} = n^{-1}\sum_{i=1}^{n} \mathbf{1}\{ \theta(\mathbf{x}_{i}) \leq 0 \}R_{i}
+ n^{-1}\sum_{i=1}^{n} \mathbf{1}\{ \theta(\mathbf{x}_{i}) > 0\}(1 - R_{i}). \nonumber
\end{equation}
Coverage proportions are measured as the average coverage over individuals,
namely, 
$n^{-1}\sum_{i=1}^{n} \mathbf{1}\{  \hat{\theta}^{L}(\mathbf{x}_{i}) \leq \theta(\mathbf{x}_{i}) \leq \hat{\theta}^{U}(\mathbf{x}_{i}) \}$,
for interval estimates $[\hat{\theta}^{L}(\mathbf{x}_{i}), \hat{\theta}^{U}(\mathbf{x}_{i})]$.

For the performance measures of RMSE and coverage proportions, 
we compared our tree-based non-parametric AFT model (NP-AFTree) with the 
semi-parametric AFT model (SP-AFTree) where the BART model is used for the regression function and
the residual distribution is assumed to be Gaussian. In addition, we compared the NP-AFTree procedure with a parametric AFT model (Param-AFT)
which assumes a linear regression with treatment-covariate interactions and log-normal residuals.

For both the NP-AFTree and SP-AFTree methods, $7,000$ MCMC iterations were used with the
first $2,000$ treated as burn-in steps. For both of these, the default parameters (i.e., $q = 0.5$, $k=2$, $J=200$)
were used for each simulation scenario. 
\vspace{-6pt}
\subsection{AFT simulations based on the SOLVD trials} \label{ss:solvd_aft}
\vspace{-3pt}
In our first set of simulations, we use data from the SOLVD trials (\cite{solvd:1991})
to guide the structure of the simulated data. Further details regarding the SOLVD
trials are discussed in Section \ref{s:hte_example}. To generate our simulated data, we first took
two random subsets of sizes $n = 200$ and $n = 1,000$ from the SOLVD data.
For each subset, we computed estimates
$\tilde{m}^{200}(A,\mathbf{x})$ and $\tilde{m}^{1000}(A, \mathbf{x})$ respectively of the regression function 
for $A \in \{0, 1\}$ using the non-parametric AFT Tree model. 
Simulated responses $y_{k}$ were then generated as
\begin{equation}
\log y_{k} = A_{k(n)}\tilde{m}^{n}(0, \mathbf{x}_{k(n)}) + (1 - A_{k(n)})\tilde{m}^{n}(1, \mathbf{x}_{k(n)}) - 0.4
+ W_{k}, \qquad k = 1, \ldots, n,
\label{eq:simulation_equation}
\end{equation}
where the regression function was fixed across simulation replications and $(A_{k(n)},\mathbf{x}_{k(n)})$
corresponds to the $k^{th}$ patient's treatment assignment and covariate vector in the random subset with $n$ patients.
The constant $-0.4$ in (\ref{eq:simulation_equation}) was added so that there was a substantial fraction of simulated
patients would have an underlying ITE $\theta(\mathbf{x}) = \tilde{m}(1, \mathbf{x}) - \tilde{m}(0, \mathbf{x}) -0.4$ less than zero.
In particular, $44.1\%$ of patients in the $n = 1,000$ simulations had an underlying value of $\theta(\mathbf{x})$ greater than zero
while $88.5\%$ had positive values of $\theta(\mathbf{x})$. For the distribution
of $W_{k}$, we considered four different choices: a Gaussian distribution, 
a Gumbel distribution with mean zero, a ``standardized" Gamma distribution with mean zero,
and a mixture of three t-distributions with $3$ degrees of freedom for each mixture component.
The parameters of each of the four distributions were chosen so that the variances were
approximately equal. The levels of censoring was varied across three levels:
none, light censoring (approximately $15\%$ of cases censored), and heavy
censoring ($45\%$ of cases censored).

Mean-squared error, classificiation (MCprop), and coverage results are shown in Figure \ref{fig:sim1_results}. 
More detailed results from this simulation study are displayed in the supplementary material.
As may be inferred from Figure \ref{fig:sim1_results}, the 
NP-AFTree method consistently performs better in terms of root mean-squared error and misclassification proportion
than the SP-AFTree procedure. Moreover, while not apparent from the figure, the NP-AFTree approach performs just as well as SP-AFTree,
even when the true residual distribution is Gaussian (see the supplementary material). 
For each residual distribution, the advantage of NP-AFTree over SP-AFTree
is more pronounced for the smaller sample sizes settings $n=200$, with closer
performance for the $n = 1,000$ cases.
While root mean-squared error and misclassification seem to be comparable between NP-AFTree and SP-AFTree
for the $n=1,000$ settings, the coverage for NP-AFTree is consistently
closer to the desired $95\%$ level and is greater than $95\%$ for nearly all
settings. When $n=200$, coverage often differs substantially from $95\%$,
but in these cases, BART is quite conservative in the sense 
that coverage is typically much greater than $95\%$.

\begin{figure}
\centering
     \includegraphics[width=3.4in,height=3.8in]{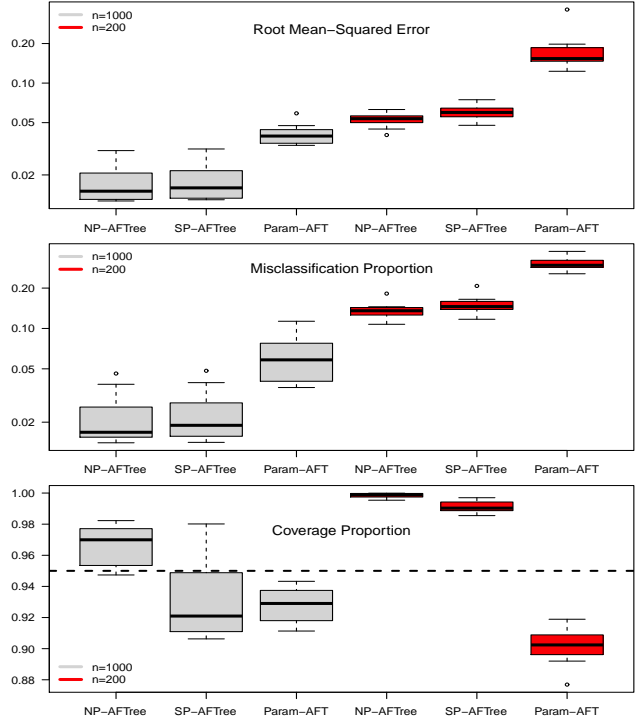}
\caption{{\footnotesize Simulations based on the SOLVD trial data.
Results are based on $50$ simulation replications. 
Root mean-squared error, Misclassification proportion, and empirical coverage
are shown for each method. Performance measures are shown for the non-parametric
tree-based AFT (NP-AFTree) method, the semi-parametric tree-based AFT (SP-AFTree),
and the parametric, linear regression - based AFT (Param-AFT) approach.
Four different choices of the residual distribution were chosen: a Gaussian distribution, 
a Gumbel distribution with mean zero, a ``standardized" Gamma distribution with mean zero,
and a mixture of three t-distributions with $3$ degrees of freedom for each mixture component.} }
\label{fig:sim1_results}
\end{figure}
\vspace{-6pt}
\subsection{Several ``Null" Simulations} \label{ss:null_sims}
\vspace{-3pt}
We considered data generated from several ``null" cases where the simulation scenarios were designed so that no HTE was present.
For these simulations, we consider data generated from four AFT models and data generated from 
a Cox proportional hazards model. In these ``null" simulations, we are primarily interested in the degree to which the NP-AFTree procedure
``detects" spurious HTE in situations where no HTE is present in the underlying data generating model.
The AFT models used for the simulations assumed a linear regression function with
no treatment-covariate interactions
\begin{equation}
\log y_{k} = \beta_{0} + \beta_{1}A_{i} + \sum_{k} \beta_{k}x_{ik} + W_{i},
\label{eq:AFT_null}
\end{equation}
and the hazard functions for the Cox model simulations similarly took the form
\begin{equation}
h(t|\mathbf{x}) = h_{0}(t)\exp\big( \beta_{0} + \beta_{1}A_{i} + \sum_{k} \beta_{k}x_{ik} \big).
\label{eq:Cox_null}
\end{equation}
Although there may be a degree of heterogeneity in $\theta(\mathbf{x})$ for the Cox proportional hazards model (\ref{eq:Cox_null}),
it is still worthwhile to investigate the behavior of $D_{i}$
when there is no HTE when treatment effects are defined in terms of hazard ratios.

The parameters in (\ref{eq:AFT_null}) were first estimated from the SOLVD data using a parametric 
AFT model with log-normal residuals. Here, we estimated the parameters in (\ref{eq:AFT_null}) separately
using the same fixed subsets of size $n = 1,000$ and $n=200$ used in Section \ref{ss:solvd_aft}.
The parameters were fixed across simulation replications.
For the AFT models, we considered the same four choices of the residual term distribution as in Section \ref{ss:solvd_aft}. 
The parameters for the hazard functions in (\ref{eq:Cox_null}) were found
by fitting a Cox proportional hazards model to the same two subsets of size $n = 200$ and $n = 1,000$ from the SOVLD trials data,
and these parameters were fixed across simulation replications.
The cumulative baseline hazard function used to generate the Cox proportional hazards simulations 
was found by using Breslow's estimator. 
In these simulations, we ran the NP-AFTree procedure with $2,000$ MCMC iterations with the
first $1,000$ treated as burn-in steps.

For each null simulation scenario, we computed the posterior probabilities of differential
treatment effect $D_{i}$ (see equations (\ref{eq:pp_dte}) and \ref{eq:pp_dte_star}) and tabulated the percentage of patients
with either strong evidence of differential treatment effect (i.e., $D_{i}^{*} > 0.95$)
or mild evidence (i.e., $D_{i}^{*} > 0.8$).
Table \ref{tab:null_sims} shows, for each null simulation scenario, the average proportion of individuals
exhibiting strong evidence of a differential treatment effect and the average proportion of individuals
exhibiting mild evidence of a differential treatment effect.
As displayed in Table \ref{tab:null_sims}, the average percentage of individuals
showing strong evidence of differential treatment effect is less than $0.22\%$ for all
simulation settings. Moreover, the percentages of cases with mild evidence of differential treatment
effect was fairly modest. The average percentage of patients with mild evidence 
was less than $3.9\%$ for all except one simulation scenario, and most of the simulation
scenarios had, on average, less than $3\%$ of patients exhibiting mild evidence of differential treatment effect.
Null simulations with $n = 200$ tended to have much fewer cases of strong or mild evidence than
those simulations with $n = 1,000$. The results presented in Table \ref{tab:null_sims}
suggest that the NP-AFTree procedure rarely reports any patients as having
strong evidence of differential treatment effects for situations where HTE is absent.

\begin{table}[ht]
\caption{{\footnotesize Simulation for settings without any HTE present.
Results are based on $100$ simulation replications.
Average \textit{percentage} of patients exhibiting strong evidence (SE) of differential treatment effect
(i.e. $D_{i} \geq 0.95$ or $D_{i} \leq 0.05$)
and average \textit{percentage} of patients exhibiting mild evidence (ME) of differential treatment effect
(i.e., $D_{i} \geq 0.8$ or $D_{i} \leq 0.2$).
Results are shown for AFT models with the same four residual distributions used in the simulations
from Section \ref{ss:solvd_aft}
and for a Cox-proportional hazards model with no treatment-covariate interactions.
Censoring levels were varied according to: none, light censoring (approximately $25\%$ of cases censored),
and heavy censoring (approximately $45\%$ of cases censored).} } 
\vspace{.3cm}
\centering
\resizebox{\textwidth}{!}{%
\begin{tabular}{rrrrrrrrrrrr}
 &  & \multicolumn{2}{c}{Normal} &
      \multicolumn{2}{c}{Gumbel} &
      \multicolumn{2}{c}{Std-Gamma} &
      \multicolumn{2}{c}{T-mixture} &
      \multicolumn{2}{c}{Cox-PH} \\
      \cmidrule(lr){3-4}\cmidrule(lr){5-6}\cmidrule(lr){7-8}
      \cmidrule(lr){9-10}\cmidrule(lr){11-12}
      n & Censoring & SE & ME & SE & ME & SE & ME & SE & ME & SE & ME \\
  \hline
    200  & none   & 0.000 & 0.095 & 0.000 & 0.040 & 0.000 & 0.070 & 0.000 & 0.060 & 0.000 & 0.140 \\ 
    200  & light  & 0.000 & 0.095 & 0.000 & 0.000 & 0.075 & 0.350 & 0.000 & 0.380 & 0.000 & 0.015 \\ 
    200  & heavy  & 0.000 & 0.020 & 0.000 & 0.000 & 0.000 & 0.000 & 0.000 & 0.055 & 0.000 & 0.290 \\ 
    1000 & none   & 0.000 & 0.803 & 0.073 & 2.066 & 0.092 & 1.483 & 0.161 & 3.087 & 0.176 & 3.824 \\ 
    1000 & light  & 0.061 & 1.989 & 0.015 & 0.967 & 0.213 & 2.674 & 0.066 & 3.176 & 0.111 & 3.405 \\ 
    1000 & heavy  & 0.002 & 1.253 & 0.000 & 0.484 & 0.030 & 1.176 & 0.123 & 2.953 & 0.135 & 2.688 \\ 
   \hline
   \hline
\end{tabular}}
\label{tab:null_sims}
\end{table}

\vspace{-6pt}
\subsection{Friedman's Randomly Generated Functions}
\vspace{-3pt}
In these simulations, we further evaluate the performance of the NP-AFTree
using randomly generated nonlinear regression functions.
To generate these random functions, we use a similar approach to that 
used in \cite{friedman:2001} to assess the performance of gradient boosted regression trees. 
This approach allows us to test our approach on a wide range of difficult nonlinear regression functions
that have higher-order interactions. 
For these simulations, we generated random regression functions $m(A, \mathbf{x})$ via
\begin{equation}
m(A, \mathbf{x}_{i}) = F_{0}(\mathbf{x}_{i}) + A_{i}\theta(\mathbf{x}_{i}),
\nonumber
\end{equation}
where the functions $F_{0}(\mathbf{x})$ and $\theta(\mathbf{x})$ are defined as
\begin{equation}
F_{0}(\mathbf{x}_{i}) = \sum_{l=1}^{10} a_{1l} g_{1l}( \mathbf{z}_{1l} )
\qquad \textrm{and} \qquad \theta( \mathbf{x}_{i} ) = \sum_{l=1}^{5} a_{2l} g_{2l}( \mathbf{z}_{2l} ).
\label{eq:friedman_fn}
\end{equation}
The coefficients in (\ref{eq:friedman_fn}) are generated as $a_{1l} \sim \textrm{Uniform}(-1,1)$
and $a_{2l} \sim \textrm{Uniform}(-0.2, 0.3)$.  
The vector $\mathbf{z}_{jl}^{i}$ is a subset of $\mathbf{x}_{i}$
of length $n_{jl}$ where the randomly selected indices used to construct the subset of $\mathbf{x}_{i}$
are the same for each $i$. The subset sizes are generated as $n_{jl} = \min(\lfloor r_{l} + 1.5 \rfloor, 10)$
where $r_{jl} \sim \textrm{Exponential}(1/2)$.
\begin{equation}
g_{jl}( \mathbf{z}_{jl} ) 
= \exp\Big\{ - \frac{1}{2} (\mathbf{z}_{jl} - \mathbf{\mu}_{jl})^{T}\mathbf{V}_{jl}( \mathbf{z}_{jl} - \mathbf{\mu}_{jl} ) \Big\}.
\nonumber
\end{equation}
The elements $\mu_{jlk}$ of the vector $\bmu_{jl}$ are generated as $\mu_{jlk} \sim \textrm{Normal}(0, 1)$,
and the random matrix $\mathbf{V}_{jl}$ is generated as 
$\mathbf{V}_{jl} = \mathbf{U}_{jl}\mathbf{D}_{jl}\mathbf{U}_{jl}^{T}$, where 
$\mathbf{D}_{jl} = \textrm{diag}\{d_{jl,1},\ldots,d_{jl,n_{jl}} \}$ with 
$\sqrt{d_{jl,k}} \sim \textrm{Uniform}(0.1,2)$ and where $\mathbf{U}_{jl}$ is a random orthogonal matrix.
We generated the covariate vectors $\mathbf{x}_{i} = (x_{i,1},\ldots,x_{i,20})^{T}$ of length $20$ independently
with $x_{i,k} \sim \textrm{Normal}(0, 1)$. Treatment assignments $A_{i}$ were generated randomly with $P(A_{i} = 1) = 1/2$.
These simulation settings imply that $\theta(\mathbf{x})$ is positive for roughly $87\%$ of individuals.
The parameters of the residual distributions were chosen so that the variances 
of each distribution were approximately equal.

\begin{figure}
\centering
     \includegraphics[width=3.4in,height=3.8in]{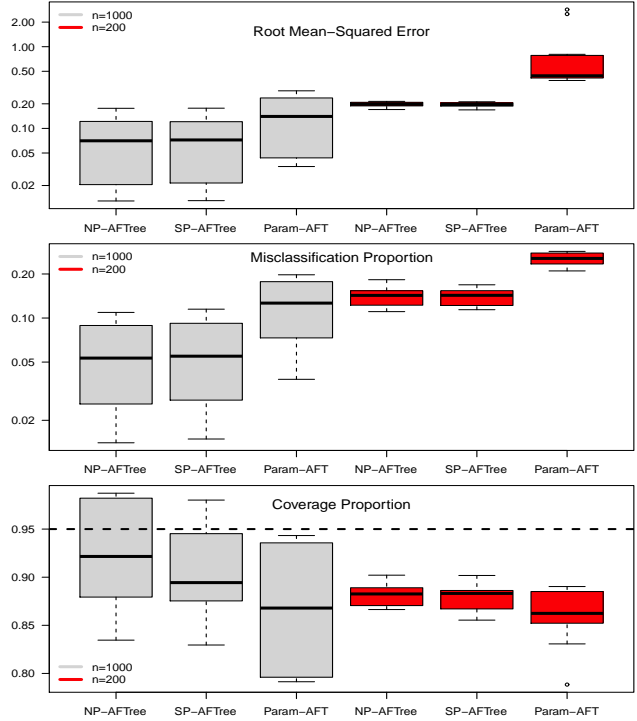}
\caption{{\footnotesize Simulations for AFT models with randomly generated regression functions. 
Results are based on $50$ simulation replications.
Root mean-squared error, misclassification proportion, and empirical coverage
are shown for each method. Performance measures are shown for the non-parametric
tree-based AFT (NP-AFTree) method, the semi-parametric tree-based AFT (SP-AFTree),
and the parametric, linear regression - based AFT (Param-AFT) approach.
Four different choices of the residual distribution were chosen: a Gaussian distribution, 
a Gumbel distribution with mean zero, a ``standardized" Gamma distribution with mean zero,
and a mixture of three t-distributions with $3$ degrees of freedom for each mixture component.} }
\label{fig:sim3_results}
\end{figure}

Figure \ref{fig:sim3_results} shows simulation results for NP-AFTree, SP-AFTree,
and the parametric AFT model. In this figure, we observe that root-mean squared
error is broadly the same for the NP-AFTree and SP-AFTree methods with
each of the tree methods exhibiting much better performance than Param-AFT.
This similarity in RMSE of SP-AFTree and NP-AFTree seems attributable to the difficulty of estimating 
these regression functions which seems to overwhelm most of the advantages of more flexible modeling of the residual distribution.
Compared to SP-AFTree, NP-AFTree shows modestly better classification performance,
particularly for settings that have non-Gaussian residual distributions and for settings with the larger ($n = 1,000$) sample size.
However, as in the simulations of Section \ref{ss:solvd_aft}, there seems
to be no advantage here of either NP-AFTree, SP-AFTree, or Param-AFT over the
naive treatment allocation approach when the sample size is $n = 200$.
These results suggest that fairly large sample sizes may be needed
for there to be any advantage over the naive approach which simply allocates
individuals to the treatment having the more beneficial overall treatment effect.
For NP-AFTree the average coverage is consistently a few percentage points
below the desired $95\%$ level suggesting that modest under-coverage
can occur in certain settings.

\vspace{-6pt}
\section{HTE in the SOLVD Trials} \label{s:hte_example}
\vspace{-3pt}
The Studies of Left Ventricular Dysfunction (SOLVD) were devised to investigate the efficacy of the 
angiotensin-converting enzyme (ACE) inhibitor 
enalapril in a target population with low left-ventricular ejection fractions.
The SOLVD treatment trial (SOLVD-T) enrolled patients determined to have a history of overt congestive
heart failure, and the SOLVD prevention trial (SOLVD-P) enrolled patients without overt congestive heart failure.
In total, $2,569$ patients were enrolled in the treatment trial while $4,228$ patients
were enrolled in the prevention trial.
The survival endpoint that we examine in our analysis is time until death or hospitalization
where time is reported in days from enrollment.

In our analysis of the SOLVD-T and SOLVD-P trials, we included $18$ patient covariates
common to both trials, in addition to using treatment and study indicators as covariates. 
Of these $18$ covariates, $8$ were continuous covariates, $8$ were binary covariates, one covariate was
categorical with three levels, and one was categorical with four levels.
In our analysis, we dropped those patients who had one or more missing covariate values,
which resulted in $548$ patients being dropped from the total of $6,797$ enrolled
in either trial. 
\vspace{-6pt}
\subsection{Cross-Validation across Hyperparameter Settings}
\vspace{-3pt}
When fitting the NP-AFT model with the SOLVD data, we considered several settings for the hyperparameters,
and for each setting of the hyperparameters, we computed cross-validation scores to
evaluate performance in terms of predicting patient outcomes and in terms of characterizing HTE.
For evaluating predictions of patient outcomes, we utilize, as in \cite{Tian:2014} and \cite{Tian:2007}, 
a direct measure of absolute prediction error. In particular, for the $k^{th}$ 
test set $\mathcal{D}_{k}$, we compute the following cross-validation score
\begin{equation}
CV_{k}^{abs} = 
\frac{1}{n_{k}}\sum_{i \in \mathcal{D}_{k} } \frac{\delta_{i} }{ \hat{V}(Y_{i}|A_{i},\mathbf{x}_{i})}
\Big| \log Y_{i} - \hat{m}_{-\mathcal{D}_{k}}(A_{i}, \mathbf{x}_{i} ) \Big|,
\label{eq:pred_cv}
\end{equation}
where $n_{k}$ is the number of patients in $\mathcal{D}_{k}$ and
$\hat{m}_{-\mathcal{D}_{k}}(A, \mathbf{x})$ is the regression function 
estimated from the $k^{th}$ training set. The weights used in (\ref{eq:pred_cv}) $\hat{V}(Y_{i}|A_{i},\mathbf{x}_{i})$
are estimates of the censoring probability $V(t|A,\mathbf{x}) = P(C > t|A, \mathbf{x})$.
The total K-fold cross-validation error is computed as $K^{-1}\sum_{k=1}^{K} CV_{k}^{abs}$.

\begin{figure}
\centering
     \includegraphics[width=4.4in,height=3.5in]{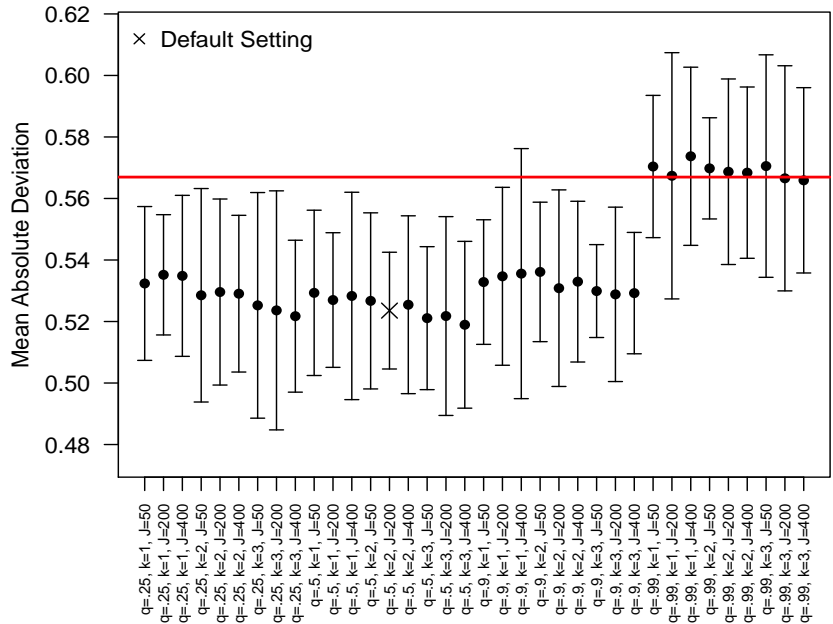}
\caption{{\footnotesize Ten-fold cross-validation for the SOLVD-T and SOLVD-P trials using the mean absolute
deviation estimate defined in (\ref{eq:pred_cv}). Twenty seven settings of the hyperparameters
are considered. The cross-validation score for the default setting of the hyperparameters
is marked with an $\times$. The horizontal red line denotes the ten-fold cross-validation
score of a parametric AFT model with log-normal errors where a linear regression
model with treatment-covariate interactions was assumed.} }
\label{fig:cross_vals}
\end{figure}

Figure \ref{fig:cross_vals} shows results from applying cross-validation to the SOLVD trials
with $36$ different settings of the hyperparameters. 
The censoring probabilities $\hat{V}(Y_{i}|A, \mathbf{x}_{i})$ used as weights in (\ref{eq:pred_cv})
were estimated using a Cox model.   
The $36$ hyperparameter settings were generated
by varying the hyperparameter $q$ which determines the parameters of the base distribution $G_{0}$, the hyperparameter
$k$ that determines the prior variance of the node values, and the number of trees $J$. 
We varied $q$ across the four levels, $q = 0.25,0.5,0.90, 0.99$; $k$ across the three levels, $k = 1,2,3$;
and the number of trees $J$ across the three levels, $J = 50, 200, 400$.  
Ten-fold cross-validation was used for each setting of the hyperparameters.
As shown in Figure \ref{fig:cross_vals}, the hyperparameter $q$ appears to play
the most important role in driving the differences in cross-validation performance
while larger values of the shrinkage parameter $k$ seem to have a modest beneficial effect in the $q = 0.25$ and $q = 0.5$ settings. 
The settings with the very conservative choice of $q=0.99$ exhibit poor performance
giving similar cross-validation scores as a parametric AFT model with an assumed linear model for the regression function.
The setting with the best cross-validation score was $q=0.5, k=3, J=400$.
This cross-validation score, however, was not notably different than many of 
the settings with either $q=0.25$ and $q = 0.5$.
For this reason, we continued to use
the default setting of $q = 0.5, k=2$, and $J=200$ in our analysis of the SOLVD trials.
\vspace{-6pt}
\subsection{Individualized Treatment Effect Estimates and Evidence for HTE}
\begin{figure}
\centering
     \includegraphics[width=4in,height=3.2in]{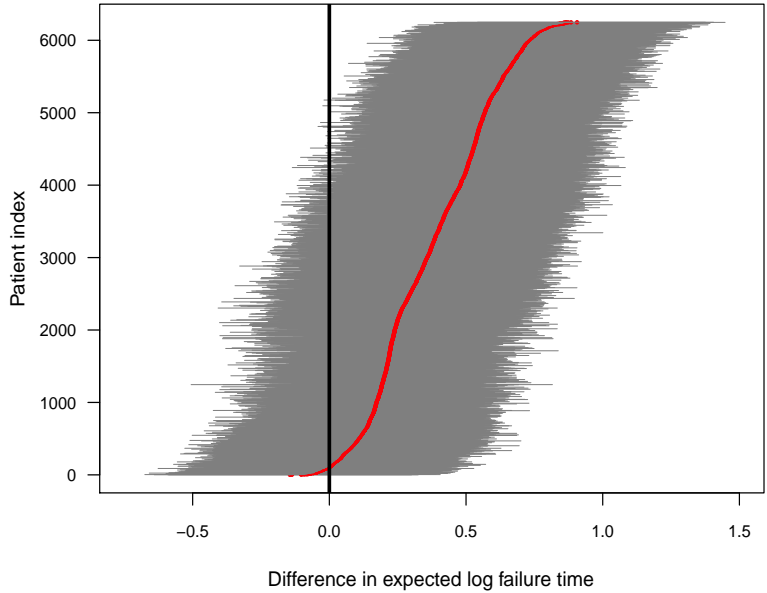}
\caption{{\footnotesize Posterior means of $\theta(\mathbf{x})$ (red points) with corresponding
$95\%$ credible intervals for patients in the SOLVD-T and SOLVD-P trials.} }
\label{fig:individualized}
\end{figure}

\vspace{-3pt}
Figure \ref{fig:individualized} shows point estimates of the ITEs $\theta(\mathbf{x})$
for patients in both the SOLVD-T and SOLVD-P trials. While the plot in Figure \ref{fig:individualized}
indicates a clear, overall benefit from the treatment, the variation in the ITEs suggests
substantial heterogeneity in response to treatment.

Examining the posterior probabilities of differential treatment effect offers
further evidence for the presence of meaningful HTE in the SOLVD trials.
Table \ref{tab:dte} shows that, in the SOLVD-T trial, approximately $19\%$
of patients had strong evidence of a differential treatment effect (i.e.
$D_{i}^{*} > 0.95$), and approximately $42\%$ of patients 
had mild evidence of a differential treatment effect (i.e., $D_{i}^{*} > 0.80$).
In the SOLVD-P trial approximately $7\%$ of patients had strong evidence of 
a differential treatment effect while approximately $32\%$ had mild evidence.
Comparison of these percentages with the results from the simulations of Section \ref{ss:null_sims} suggests
the presence of HTE. In the null simulation scenarios of Section \ref{ss:null_sims},
the proportion of cases with strong evidence of differential treatment was very close
to zero. Thus, the large proportion of patients with strong evidence 
for differential treatment effect is an indication of the presence of HTE
in the SOLVD trials that deserves further exploration.

\vspace{-6pt}
\subsection{Characterizing Variation in Treatment Effect}
\vspace{-3pt}
Figure \ref{fig:histogram_ests} displays a histogram of the posterior means of the treatment ratios
$\xi( \mathbf{x} ) = E( T | A=1, \mathbf{x}, m )/E(T | A = 0, \mathbf{x}, m )$,
for each patient in the SOLVD-T and SOLVD-P trials. 
In contrast to the ITE scale used in Figure \ref{fig:individualized},
defining the ITEs in terms of the ratios of expected failure times may provide a more interpretable scale by which to describe HTE.
As may be inferred from the histogram in Figure \ref{fig:histogram_ests}, nearly all patients have a
positive estimated treatment effect with $98.9\%$ having an estimated value of $\xi(\mathbf{x}_{i})$
greater than one. Of those in the SOLVD-T trial, all the patients had 
$E\{ \xi(\mathbf{x}_{i}) | \mathbf{y}, \bdelta  \} > 1$,
and $98.2\%$ of patients in the SOLVD-P trial had $E\{ \xi(\mathbf{x}_{i}) |\mathbf{y}, \bdelta \} > 1$.

Figure \ref{fig:histogram_ests} also reports the smoothed estimate $\hat{h}_{n}(t)$ of the distribution of the treatment effects separately
for the two trials. These smoothed posterior estimates of the treatment effect distribution 
were computed as described in equation (\ref{eq:smooth_density}) where
posterior samples of $\xi(\mathbf{x}_{i})$ were used in place of $\theta(\mathbf{x}_{i})$.
Note that the $\hat{h}_{n}( t )$ shown in Figure \ref{fig:histogram_ests} are estimates of the distribution
of the underlying treatment effects and do not represent the posterior distribution of the overall
treatment effects within each trial.
As expected, the variation in treatment effect suggested by the plots of $\hat{h}_{n}(t)$ in Figure \ref{fig:histogram_ests}
is greater than the variation exhibited by the posterior means of $\xi(\mathbf{x}_{i})$. 
The estimates $\hat{h}_{n}(t)$ provide informative characterizations of the distribution of treatment
effects in each trial especially for visualizing the variability in treatment effects in each trial.
The posterior median of the standard deviation of treatment effect 
$\sqrt{ \sum_{i} \{\xi(\mathbf{x}_{i}) - \bar{\xi} \} ^{2}}$ was 
$0.45$ for the SOLVD-T and $0.34$ for the SOLVD-P trial.
\vspace{-6pt}
\subsection{Proportion Benefiting}
\begin{figure}
\centering
     \includegraphics[width=5in,height=4.1in]{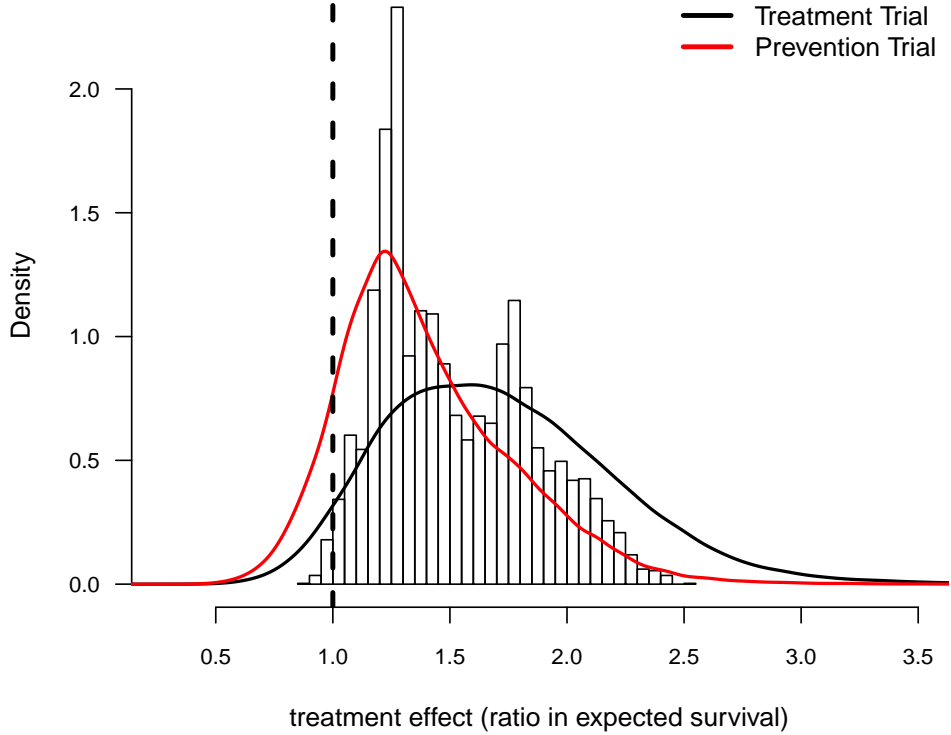}
\caption{{\footnotesize Histogram of point estimates (i.e., posterior means) of the treatment effects 
$\xi(\mathbf{x}) = e^{\theta(\mathbf{x})}$
and smooth posterior estimates $\hat{h}_{n}(t)$ of the treatment effect distribution.  
The histogram is constructed using all point estimates from both the SOLVD treatment and prevention trials.
Smooth estimates, $\hat{h}_{n}(t)$ , of the distribution of treatment effects were computed as described
in equation for the two trials separately. The kernel bandwidth $\lambda$ for each trial was chosen 
using the rule $\lambda = [0.9 \times \min(\hat{\sigma}_{\xi}, \hat{IQR}_{\xi})]/[1.34 \times n_{t}^{1/5}]$, where
$\hat{\sigma}_{\xi}$ and $\hat{IQR}_{\xi}$ are posterior means of the standard deviation and inter-quartile range of 
$\xi(\mathbf{x}_{i})$ respectively and where $n_{t}$ is the trial-specific sample size.} }
\label{fig:histogram_ests}
\end{figure}

\vspace{-3pt}
We can use (\ref{eq:prop_benefit}) to estimate the proportion of patients benefiting
in each trial. The estimated proportions of patients (i.e., the posterior mean of $Q$ in (\ref{eq:prop_benefit}))
benefiting were $95.6\%$ and $89.1\%$ in the SOLVD-T and SOLVD-P trials respectively. 
These proportions are approximately equal to the area under the curve of $\hat{h}_{n}(t)$ for $t \geq 1$
in Figure \ref{fig:histogram_ests}.

Table \ref{tab:dte} shows a tabulation of patients according to evidence of treatment benefit.
In both trials, all patients have at least a $0.25$ posterior probability of treatment benefit
(i.e. $P\{ \xi(\mathbf{x}_{i}) > 1| \mathbf{y}, \bdelta \} > 0.25$).
In the treatment trial, $76\%$ percent of patients exhibit a posterior probability of benefit
greater than $0.95$, and the percentage 
for the prevention trial is $44\%$.

\begin{table}[ht]
\centering
\begin{tabular}{rrr}
  \hline
 & SOLVD Treatment Trial & SOLVD Prevention Trial \\ 
  \hline
 $P\{ \xi(\mathbf{x}_{i}) > 1| \mathbf{y}, \bdelta \} \in (0.99,1]$         & 51.38 & 20.47 \\ 
  $P\{ \xi(\mathbf{x}_{i}) > 1| \mathbf{y}, \bdelta \} \in (0.95,0.99]$      & 24.69 & 23.71 \\ 
  $P\{ \xi(\mathbf{x}_{i}) > 1| \mathbf{y}, \bdelta \} \in (0.75,0.95]$      & 20.08 & 41.98 \\ 
  $P\{ \xi(\mathbf{x}_{i}) > 1| \mathbf{y}, \bdelta \} \in (0.25,0.75]$      & 3.85 & 13.84 \\ 
   $P\{ \xi(\mathbf{x}_{i}) > 1| \mathbf{y}, \bdelta \} \in [0,0.25]$        & 0.00 & 0.00 \\ 
  \hline
  \hline
\rule{0pt}{2ex}    
  $D_{i}^{*} > 0.95$ & 19.36 & 7.30 \\ 
  $D_{i}^{*} > 0.80$ & 41.93 & 31.58 \\ 
   \hline
\end{tabular}
\caption{{\footnotesize Tabulation of posterior probabilities of treatment benefit and posterior probablities
of differential treatment effect $D_{i} = P\{ \xi(\mathbf{x}_{i}) \geq \bar{\xi}| \mathbf{y}, \bdelta \}$. 
For each trial, the empirical percentage of patients whose estimated posterior probability of treatment benefit 
lies within each of the intervals $(0.99,1], (0.95,0.99], (0.75,0.95], (0.25,0.75]$,
and $[0,0.25]$ is reported.
In addition, the percentages of patients in each trial that exhibit ``strong" 
(i.e., $D_{i}^{*} > .95$)
and ``mild" (i.e., $D_{i} > 0.80$)
evidence of differential treatment effect are shown. }} 
\label{tab:dte}
\end{table}

\vspace{-6pt}
\subsection{Individual-level Survival Functions}
\begin{figure}
\centering
     \includegraphics[width=4.7in,height=3.5in]{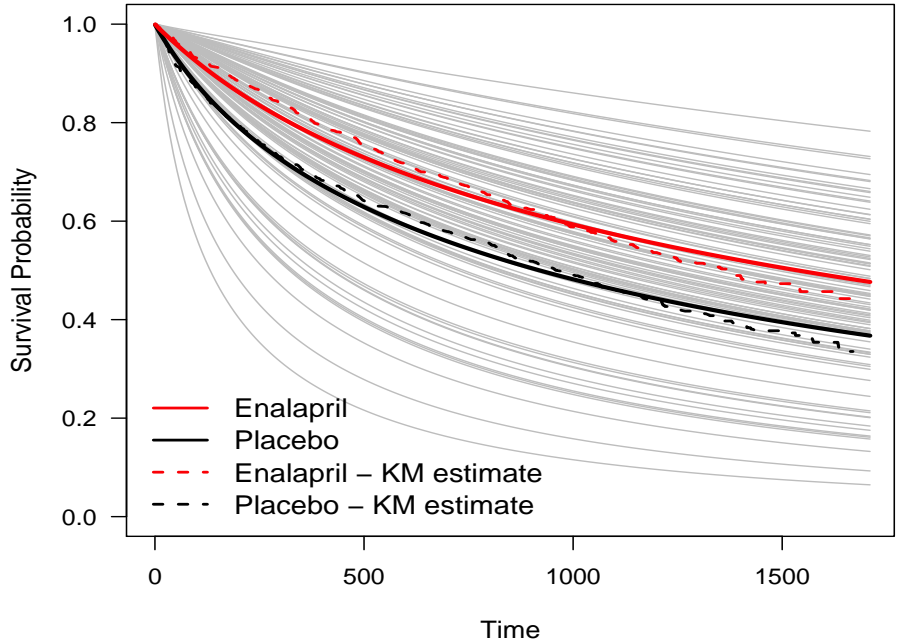}
\caption{{\footnotesize Estimates of individual-specific survival curves for selected patients from the SOLVD treatment trial.
For each patient, the posterior mean of the survival functions $P\{ T > t|A, \mathbf{x}, m, G_{H}, \sigma \}$
as defined in (\ref{eq:survival_formula}) are plotted. 
The solid black and red survival curves are the average by treatment group of these estimated individual-specific
surves. The dashed survival curves are the Kaplan-Meier estimates for each treatment group.} }
\label{fig:survival_curves}
\end{figure}

\vspace{-3pt}
Figure \ref{fig:survival_curves} shows estimated survival curves for randomly selected patients from
the SOLVD treatment trial. Averages of these individual-level survival curves
are computed for each treatment arm and compared with the corresponding Kaplan-Meier
estimates of survival. It is apparent from Figure \ref{fig:survival_curves} that considerable
heterogeneity in patient risk is present. Indeed, in the control arm, $20\%$ percent of patients
had an estimated median survival time less than $500$ days, $54\%$ had between $500$ and $1500$
days, and $26\%$ had an estimated median survival time of more than $1500$ days.

\vspace{-6pt}
\subsection{Exploring Important Variables for HTE}
\vspace{-3pt}
To explore the patient attributes important in driving differences in treatment effect,
we use a direct approach similar to the ``Virtual Twins" method used by \cite{Foster:2011}
in the context of subgroup identification. In \cite{Foster:2011}, the authors suggest a two-stage
procedure where one first estimates treatment difference for each individual and then, using these estimated
differences as a new response variable, one estimates a regression model in order to identify a region
of the covariate space where there is an enhanced treatment effect.
Similarly, to examine important HTE variables, we first fit the full non-parametric AFT model
to generate posterior means $\hat{\theta}(\mathbf{x}_{i})$ of the individualized treatment effect for each patient.
Then, we estimate a regression using the previously estimated $\hat{\theta}(\mathbf{x})$ as the response
variable and the patient covariates (except for treatment assignment) as the predictors.
Because the treatment difference $\theta(\mathbf{x})$
should only depend on covariates that are predictive of HTE, using the unobserved $\theta(\mathbf{x}_{i})$
as the responses in a regression with the patient covariates as predictors represents
a direct and efficient approach to exploring variables involved in driving treatment effect heterogeneity.

\begin{figure}
\centering
     \includegraphics[width=4.9in,height=3.6in]{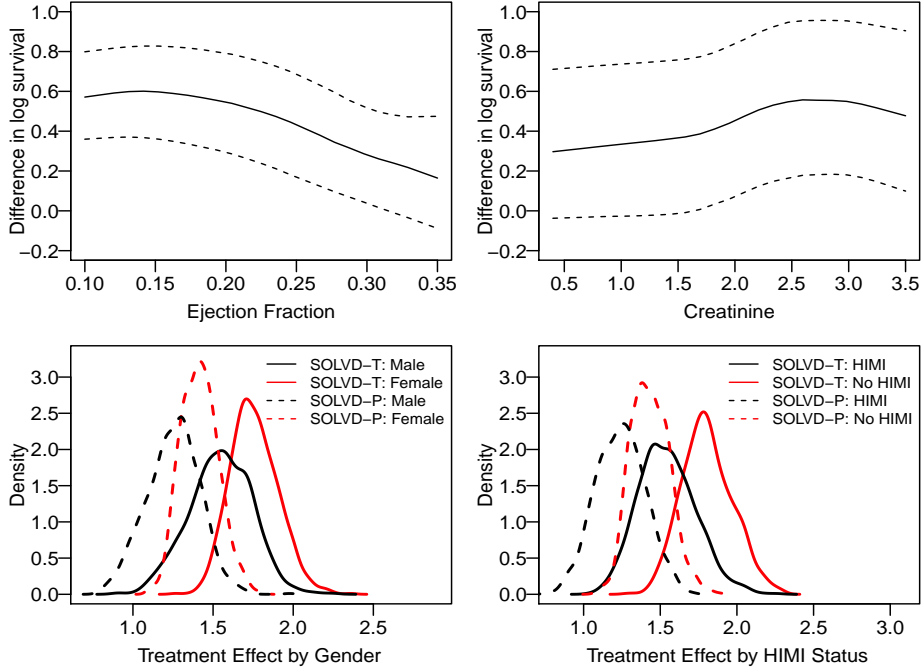}
\caption{{\footnotesize Smoothed partial dependence plots for ejection fraction and creatinine levels, 
and posterior distributions of treatment effect for men vs. women and for those
with a history of myorcardial infarction vs. those with no history of myocardial 
infarction (HIMI vs. No HIMI).} }
\label{fig:pd_plots}
\end{figure}

To investigate the important HTE variables in the SOLVD trials using the virtual twins approach, 
we fit a linear regression using weighted least squares with
$\hat{\theta}(\mathbf{x}_{i})$ as the responses and where
the residual variances were assumed proportional to the posterior variances of $\theta(\mathbf{x}_{i})$.
In this weighted regression, all covariates were normalized to have zero mean and unit variance. 
The patient covariates with the five largest estimated coefficients in absolute value
were as follows: baseline ejection fraction, history of myocardial infarction,
baseline creatinine levels, gender, and diabetic status. 
Figure \ref{fig:pd_plots} displays partial dependence plots for ejection fraction and creatinine
along with the posterior distribution of the average treatment effect in the male/female groups
and the subgroups defined by history of myocardial infarction.
The partial dependence plots clearly demonstrate an enhanced treatment effect for those
patients with lower baseline ejection fraction. The strongest change in treatment effect occurs in the ranges 
$0.20 - 0.30$.
The comparisons according to gender shown in Figure \ref{fig:pd_plots} 
show greater treatment benefit in men vs. women for both SOLVD trials.

\vspace{-6pt}
\section{Discussion} \label{s:discussion}
\vspace{-3pt}
In this paper, we have described a flexible, tree-based approach to examining heterogeneity of treatment 
effect with survival endpoints. This method produces estimates of individualized treatment effects
with corresponding credible intervals for AFT models with an arbitrary regression function
and residual distribution. Moreover, we have demonstrated how this framework
provides a useful framework for addressing a variety of other HTE-related questions.
When using the default hyperparameter settings, 
the method only requires the user to input the survival outcomes, 
treatment assignments, and patient covariates. As shown in several simulation studies,
the default settings exhibit strong predictive performance and good coverage properties. 
Though quite flexible, our non-parametric AFT model  
does entail some assumptions regarding the manner in which the patient covariates modify the baseline hazard.
Hence, it would be worth further investigating the robustness of the nonparametric AFT method to other forms of model misspecification
such as cases where neither an AFT or a Cox proportional hazards assumption holds or
cases where the residual distribution depends on the patient covariates.

In addition to describing a novel non-parametric AFT model, we examined a number of measures for reporting
HTE including the distribution of individualized treatment effects, the proportion of patients benefiting from treatment,
and the posterior probabilities of differential treatment effect. 
Each have potential uses in allowing for more refined interpretations of clinical trial results.
The argument has been made by some (e.g., \cite{kent:2007}) that the positive results
of some clinical trials are driven substantially by the outcomes of high-risk patients.
In such cases, the posterior distribution of the ITEs along with the estimated proportion
benefiting may help in clarifying the degree to which lower risk patients 
are expected to benefit from the proposed treatment. In Section \ref{ss:null_sims}, 
we explored the use of posterior probabilities of differential treatment effect 
as a means of detecting the presence of HTE. Such measures show potential 
for evaluating the consistency of treatment and for assessing
whether or not further investigations into HTE are warranted.

As described in this paper, the BART-based nonparametric AFT model only works when no missing values of 
the patient covariates occur. Though not explored in this paper, tree-based methods
have the potential to provide intuitive, automatic approaches for handling missing data.
In the ``Missing in Attributes" approach discussed by \cite{Twala:2008} and by \cite{Kapelner:2015} for BART,
the splitting rules are directly constructed to account for possible missingness.
Such an approach could potentially be incorporated into the BART framework and
allow one to handle missing covariates without needing to specify a method for imputation.

In the AFT model discussed in Sections \ref{ss:notation} and \ref{ss:cdp}, the distribution of the residual term is assumed to be the same for
all values of the covariates. Greater flexibility may be gained by relaxing this assumption by
considering an additional ``heteroskedastic" AFT model which allows
the residual distribution to change with the covariates. 
Although we do not fully explore the use of a heteroskedastic model in this work,
one approach for modifying the nonparametric AFT model
would be to allow the scale parameter of the residual density to depend on the patient covariates
while having the mixing distribution remain independent of these covariates.
One direct way of allowing the scale parameter to vary with the covariates
would be to use a parametric log-linear model as described in 
the approach of \cite{Bleich:2014}.

We note that our approach models the regression function $m(A, \mathbf{x})$ for all values of $A$ and $\mathbf{x}$ 
in order to compute treatment effect contrasts $m(1, \mathbf{x}) - m(0, \mathbf{x})$ and as such, requires
modeling of both the ``main effects" and the treatment-covariate interactions. 
This general approach to analyzing HTE - referred to as ``global outcome modeling" by \cite{lipkovich:2016} -
may be contrasted with other approaches that only seek to directly model the treatment effects
without any consideration of the main effects.  
Modified outcome methods such as those described in \cite{Tian_jasa:2014} and \cite{Weisberg:2015}
analyze modified response variables whose expectations have the desired treatment effects, and
this approach allows one to directly estimate treatment-covariate interaction effects without
having to model the main effects.  
Adopting an approach similar to this would be straightforward with our accelerated failure time model.
One would only need to create modified responses by multiplying the observed log-failure times with 
an appropriate function of treatment assignment. These new responses would now exhibit
different form of censoring, but this could easily be incorporated into our data augmentation procedure used in posterior sampling.

\appendix

\def\spacingset#1{\renewcommand{\baselinestretch}%
{#1}\small\normalsize} \spacingset{1}

\spacingset{1.45}

\section{Approximate Distribution of the Residual Variance}

As discussed in Section $2.4$ of the main paper, the variance of the residual term
may be expressed as
\begin{equation}
\var(W|G, \sigma) = \sigma^{2} + \sigma_{\tau}^{2}\sum_{h=1}^{\infty}\frac{\pi_{h}}{\sigma_{\tau}^{2}}(\tau_{h}^{*} - \mu_{G^{*}})^{2}
\end{equation}
When assuming (as we do) that $\sigma_{\tau}^{2} = \kappa$, this becomes
\begin{equation}
\var(W|G, \sigma) = \sigma_{\tau}^{2}\Big[
\sigma^{2}/\kappa + \sum_{h=1}^{\infty}\frac{\pi_{h}}{\sigma_{\tau}^{2}}(\tau_{h}^{*} - \mu_{G^{*}})^{2}\Big]
\end{equation}
Because we assume that $G \sim CDP(M, G_{0})$ with $G_{0}$ as a $\text{Normal}(0, \sigma_{\tau}^{2})$
distribution, the term $[(\tau_{h}^{*} - \mu_{G^{*}})^{2}]/\sigma_{\tau}^{2}$
has a standard normal distribution.

\begin{figure}
\centering
     \includegraphics[width=6in,height=5in]{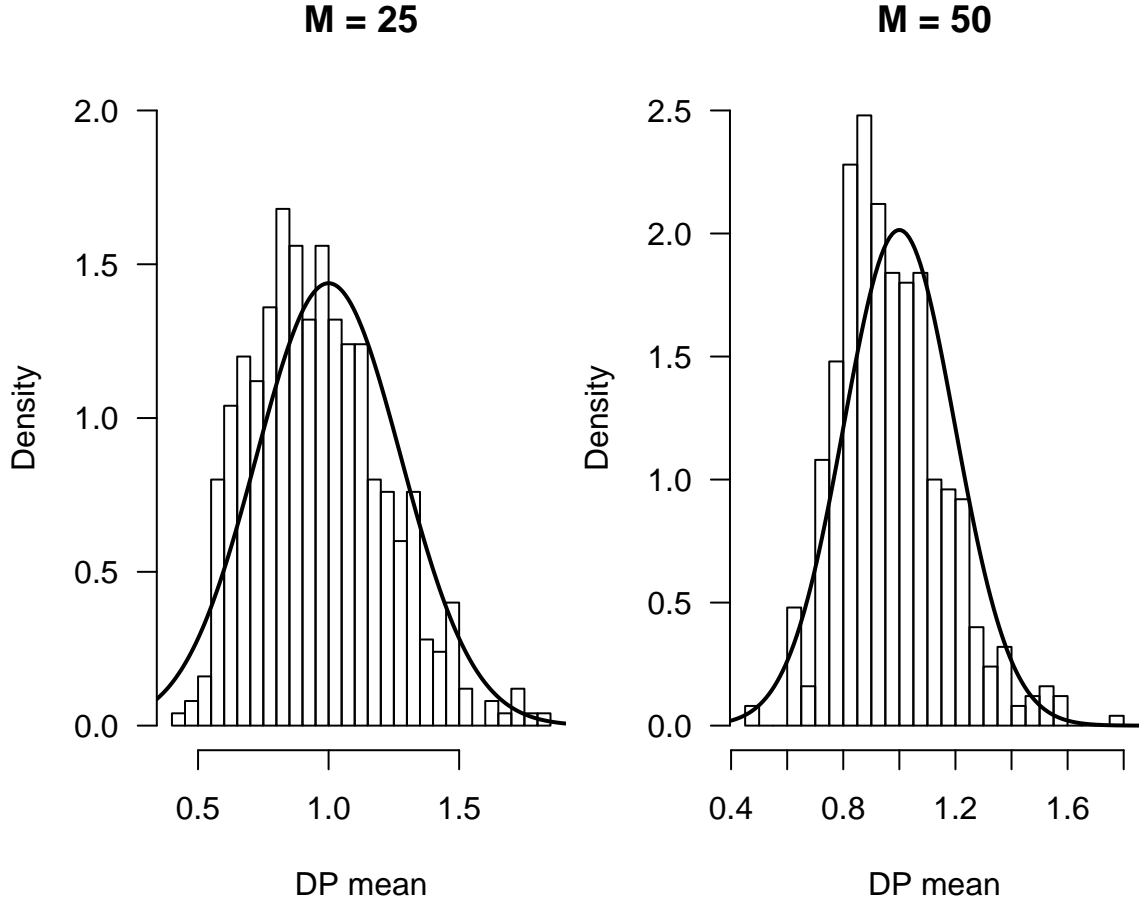}
\caption{Histogram of simulated values of $\sum_{h=1}^{\infty}\frac{\pi_{h}}{\sigma_{\tau}^{2}}(\tau_{h}^{*} - \mu_{G^{*}})^{2}$
along with the approximating $\text{Normal}\{ 1, 2/(M + 1) \}$ density.
Simulations were performed with $M = 25$ and $M = 50$ for the mass parameter.
In each case, $500$ simulated values of $a$ were drawn.}
\label{fig:DPmean}
\end{figure}

In Section $2.4$ of the main paper, it is stated that the prior distribution of $\var(W|G,\sigma)$
is approximated with the following distribution 
\begin{equation}
\sigma_{\tau}^{2}\big[ \nu/\chi_{\nu}^{2} + \text{Normal}(1, \{2(M+1)\}^{-1}) \big]. 
\label{eq:approx_prior}
\end{equation}
The above approximation relies on the fact that 
$\sum_{h=1}^{\infty}\frac{\pi_{h}}{\sigma_{\tau}^{2}}(\tau_{h}^{*} - \mu_{G^{*}})^{2}$
has an approximate $\text{Normal}(0, \{2(M+1)\}^{-1})$ distribution in the sense 
described by \cite{Yamato:1984}.
A histogram of simulated values of
$\sum_{h=1}^{\infty}\frac{\pi_{h}}{\sigma_{\tau}^{2}}(\tau_{h}^{*} - \mu_{G^{*}})^{2}$
along with a plot of the approximating
$\text{Normal}(0, \{2(M+1)\}^{-1})$
density is shown in Figure \ref{fig:DPmean}. In this figure, histograms are shown for the cases of $M = 25$ and $M = 50$.

\begin{figure}
\centering
     \includegraphics[width=6in,height=5in]{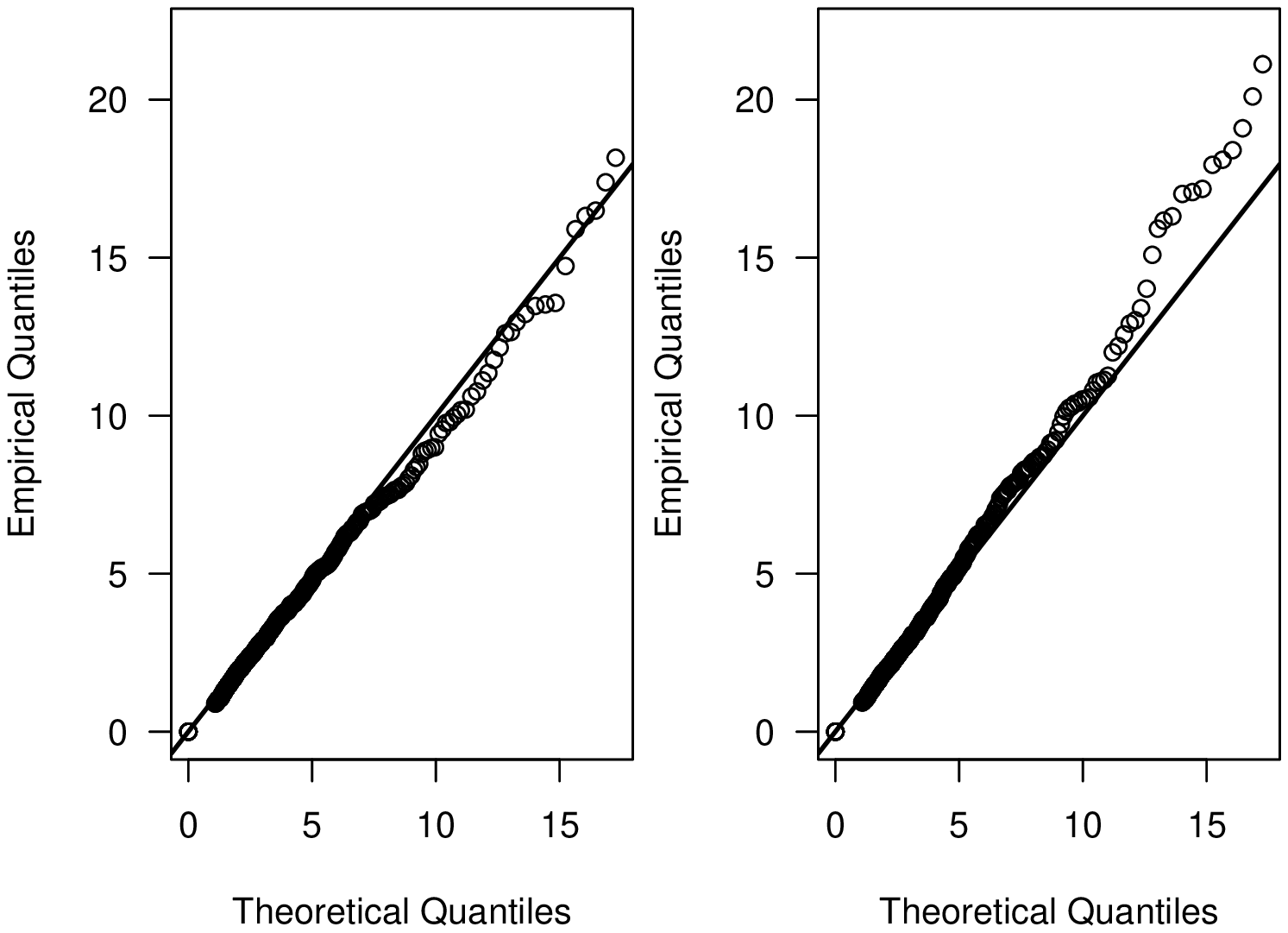}
\caption{Quantile-Quantile plot with simulated values of $\var(W|G,\sigma)$
(using equation). }
\label{fig:variance_qq}
\end{figure}

In Figure \ref{fig:variance_qq}, we display a quantile-quantile plot of simulated values from the distribution of $\var(W|G,\sigma)$
vs. the approximate theoretical quantiles obtained from the approximate prior
distribution stated in (\ref{eq:approx_prior}).

\bibliographystyle{Chicago}
\bibliography{aft_references}
\end{document}